\documentclass[runningheads]{llncs}
\usepackage{graphicx}
\usepackage{hyperref}
\usepackage{subfigure}
\usepackage{amsmath}
\usepackage{amssymb}
\usepackage{amsfonts}
\usepackage{multirow}
\usepackage{float}
\usepackage{mathtools,booktabs,tabu,graphicx,hyperref,epstopdf,url}

\usepackage[noend]{algpseudocode}

\usepackage{algorithm}
\usepackage{fancyref}
\usepackage[noend]{algpseudocode}

\usepackage[compatibility=false]{caption}% http://ctan.org/pkg/caption
\usepackage{pifont}
\usepackage[noadjust]{cite}

\usepackage[dvipsnames]{xcolor}
\usepackage{color}
\usepackage[color,matrix,arrow]{xy}

\newcommand{\G}{\mathcal{G}}

\newcommand{\ey}{\textbf{\emph{\emph{y}}}}

\newcommand{\eg}{\textbf{\emph{\emph{g}}}}

\newcommand{\ex}{\textbf{\emph{\emph{x}}}}

\newcommand{\eH}{\textbf{\emph{\emph{H}}}}

\newcommand{\emm}{\textbf{\emph{\emph{m}}}}

\newcommand{\eC}{\mathbb{C}}

\newcommand{\BLC}{\textsc{Bloch}}

\newcommand{\proxnet}{\textsc{PGD-Net}}
\newcommand{\proxnetB}{\textsc{ProxNet }}

\begin{document}
\title{Compressive MR Fingerprinting reconstruction with Neural Proximal Gradient iterations}

\titlerunning{Deep proximal compressive MRF reconstruction}

\author{Dongdong Chen\inst{1}\and Mike E. Davies\inst{1}\and Mohammad Golbabaee\inst{2}}
\authorrunning{D. Chen et al.}

\institute{School of Engineering, University of Edinburgh, UK \and Computer Science department, University of Bath, UK\\\email{\{d.chen, mike.davies\}@ed.ac.uk}, \email{mg2105@bath.ac.uk}}%\

\maketitle              % typeset the header of the contribution

\begin{abstract}
Consistency of the predictions with respect to the physical forward model is pivotal for reliably solving inverse problems. This consistency is mostly un-controlled in the current end-to-end deep learning methodologies proposed for the Magnetic Resonance Fingerprinting (MRF) problem. To address this, we propose \proxnet, a learned proximal gradient descent framework that directly incorporates the forward acquisition and Bloch dynamic models within a recurrent learning mechanism. The  \proxnet\ adopts a compact neural proximal model for de-aliasing and quantitative inference, that can be flexibly trained on scarce MRF training datasets. Our numerical experiments show that the \proxnet\ can achieve a superior quantitative inference accuracy, much smaller storage requirement, and a comparable runtime to the recent deep learning MRF baselines, while being much faster than the dictionary matching schemes. Code has been released at \url{https://github.com/edongdongchen/PGD-Net}.

\end{abstract}

\section{Introduction}
Magnetic resonance fingerprinting (MRF) is an emerging technology that enables simultaneous quantification of multitudes of tissues' physical properties in short and clinically feasible scan times~\cite{ma2013magnetic}. Iterative reconstruction methods based on Compressed Sensing (CS)  have proven efficient to help MRF overcome the challenge of computing accurate quantitative images from the undersampled k-space measurements taken in aggressively short scan times~\cite{davies2014compressed, asslander2018low, zhao-LR-MRF, doneva-LRMRF}. However, these methods require dictionary matching (DM) that is non-scalable and can create enormous storage and computational overhead. Further, such approaches often do not fully account for the joint spatiotemporal structures of the MRF data which can lead to poor reconstructions~\cite{golbabaee2020compressive}.

Deep learning methodologies have emerged to address DM's computational bottleneck~\cite{cohen-DRONE, golbabaee2019geometry, lustig-deepMRF, MRFRNN}, and in some cases to perform joint spatiotemporal MRF processing through using convolutional layers~\cite{hoppe2017deep,fang2019deep,chen2019deep, balsigerMRFDL, hoppe2019rinq, song2019hydra, fang2019rca}.
These models are trained in an end-to-end fashion \emph{without} an explicit account for the known physical acquisition model (i.e. the forward operator) and a mechanism for explicitly enforcing measurement consistency according to this sampling model which can be crucial in the safety-first medical applications. Further, ignoring the structure of the forward model could lead to building unnecessary large inference models and possible overfitted predictions, especially for the extremely \emph{scarce} labelled anatomical quantitative MRI datasets that are available for training.

\textit{Our contributions:}
we propose \proxnet\ a deep convolutional model that is able to learn and perform robust spatiotemporal MRF processing, and  work with limited access to the ground-truth (i.e. labelled) quantitative maps. Inspired by iterative proximal gradient descent  (PGD) methods for CS reconstruction~\cite{mardani2018neural}, we adopt learnable, compact and shared convolutional layers within a data-driven proximal step, meanwhile explicitly incorporating the acquisition model as a non-trainable gradient step in all iterations. The proximal operator is an auto-encoder network whose decoder embeds the Bloch magnetic responses and its convolutional encoder embeds a de-aliasing projector to the tissue maps' quantitative properties. Our work is inspired by recent general  CS methodologies~\cite{mardani2018neural, rick2017one, adler2017solving, aggarwal2018modl, chen2019decomposition} that replace traditional hand-crafted image priors by deep data-driven models. To the best of our knowledge, this is the first work to adopt and investigate the feasibility of such an approach for solving the MRF inverse problem.

\section{Methodology}
MRF adopts a linear spatiotemporal compressive acquisition model:
\begin{equation}\label{eq:sampling}
    \ey=\overline{\eH}(\overline{\ex}) +\xi
\end{equation}
where $\ey\in \eC^{m\times L}$ are the k-space measurements collected
at $L$ temporal frames and corrupted by some noise $\xi$, and $\overline \ex\in \eC^{n\times L}$ is the Time-Series of Magnetisation Images (TSMI) with $n$ voxels across $L$ timeframes.  The forward operator $\overline \eH:\mathbb{C}^{n\times L}\rightarrow \eC^{m\times L}$ models Fourier transformations subsampled according to a set of temporally-varying k-space locations in each timeframe. Accelerated MRF acquisition implies working with heavily under-sampled data $m\ll n$, which makes $\overline \eH$ becomes ill-posed for the inversion. % $\Omega$.
\newline
\textbf{Bloch response model:} Per-voxel TSMI temporal signal evolution is related to the quantitative NMR parameters/properties such as $\{T1_v,T2_v\}$ relaxation times, through the solutions of the \emph{Bloch differential equations}  $\overline \ex_v\approx \rho_v \overline{\mathcal{B}}(T1_v,T2_v)$, scaled by the $\rho_v$ proton density (PD) in each voxel $v$~\cite{ma2013magnetic,jiang2015mr}.
%\footnote{$\ex_v$ denotes the $v$th row (i.e. a multi-dimensional voxel) of the matrix $\ex$.}
\newline
\textbf{The subspace dimension-reducing model:} In many MRF applications (including ours) a low $s\ll L$ dimensional subspace $V\in \mathbb{C}^{L\times s}$ embeds the Bloch solutions $\mathcal{B}(.)\approx VV^H\mathcal{B}(.)$. This subspace can be computed through PCA decomposition of the MRF dictionary \cite{mcgivney2014svd}, and enables re-writing \eqref{eq:sampling} in a compact form that is beneficial to the storage, runtime and accuracy of the reconstruction~\cite{asslander2018low, golbabaee2019geometry}:
\begin{equation}\label{eq:sampling2}
    \ey=\eH(\ex) +\xi \,\,\,\,\, \text{where, for each voxel} \,\,\,\, \ex_v \approx \rho_v \mathcal{B}(T1_v,T2_v),
\end{equation}
and $\ex\in\eC^{n\times s}$ is the dimension-reduced TSMI, $\eH := \overline{\eH}\circ V$ and $\mathcal{B} =V^H \overline{\mathcal{B}}$ denotes the subspace-compressed Bloch solutions (for more details see~\cite{golbabaee2020compressive}).

\textbf{Tissue quantification:} Given the compressed measurements $\ey$,
the goal of MRF is to solve the inverse problem~\eqref{eq:sampling2} and to compute the underlying multi-parametric maps $\emm = \{T1, T2, \rho\}$ (and $\ex$ as a bi-product).
Such problems are typically casted as an optimisation problem of the form:
    \begin{equation}\label{eqs:pnp_regularization}%\label{loss:inverse}
     \arg\min_{\ex,\emm}\|\ey-\eH\ex\|_2^2 + \phi(\ex,\emm),
    \end{equation}
 and solved iteratively by the proximal gradient descent (PGD):
\begin{equation}\label{eq:PDG}
\text{PGD}: \left\{
\begin{array}{ll}
\eg^{(t+1)} = \ex^{(t)} + \alpha^{(t)}\eH^{H}(\ey- \eH\ex^{(t)})  & \rightarrow\hbox{\textcolor{RoyalBlue}{Gradient with step size $\alpha$}}\\
\{\ex^{(t+1)},\emm^{(t+1)}\} = \text{Prox}_{\phi}(\eg^{(t+1)})& \rightarrow\hbox{\textcolor{RoyalBlue}{Proximal update}}
\end{array}
\right.
\end{equation}
where the gradient updates encourage k-space fidelity (the first term of \eqref{eqs:pnp_regularization}), and the \emph{proximal operator} $\text{Prox}_{\phi}(\cdot)$ enforces image structure priors through a \emph{regularisation} term $\phi(\cdot)$  that makes the inverse problem well-posed.
The Bloch dynamics in \eqref{eq:sampling2} place an important temporal constraint (prior) for per-voxel trajectories of $\ex$. Projecting onto this model (i.e. a temporal Prox model) has been suggested via iterative dictionary search schemes~\cite{davies2014compressed, asslander2018low}.
This approach boost MRF reconstruction accuracy compared to the non-iterative DM \cite{ma2013magnetic}, however, DM is non-scalable and can create enormous storage and computational overhead. Further, such approach processes data independently per voxel and neglects important spatial domain regularities in the TSMIs and quantitative maps.

\section{PGD-Net for MRF quantification}%\vspace{-.2cm}
%Inspired by \cite{mardani2018neural},
We propose to learn a data-driven proximal operator within the PGD mechanism for solving the MRF problem.  Implemented by compact networks with convolutional layers, the \emph{neural} Prox improves the storage overhead and the sluggish runtime of the DM-based PGD by orders of magnitudes. Further, trained on quantitative MR images, the neural Prox network learns to simultaneously enforce spatial- and temporal-domain data structures within PGD iterations.

\textbf{Prox auto-encoder:} We implement $\text{Prox}: \eg\rightarrow \{\ex,\emm\}$ through a deep convolutional auto-encoder network:
 \begin{equation}\label{eqs:proxae}
    %\text{Prox}:=\BLC \circ \G, %\triangleq
    \text{Prox}:=\BLC \circ \G, %\triangleq
\end{equation}
consisting of an encoder $\G: \eg \rightarrow \emm$ and a decoder \BLC: $\emm \rightarrow \ex$ subnetworks. The \emph{information bottleneck} in the (neural) Prox auto-encoder corresponds to projecting multichannel TSMIs to the low-dimensional manifold of the tissues' intrinsic  (quantitative) property maps~\cite{golbabaee2020compressive}.

\textbf{Decoder network:} creates a differentiable model for \emph{generating} the Bloch magnetic responses. This network uses $1\times1$ filters to process image time-series in a voxel-wise manner.
Given quantitative properties $\emm_v=\{T1_v,T2_v,\gamma_v\}$, the decoder \emph{approximates} (dimension-reduced) Bloch responses in voxel $v$ i.e.
 $\BLC(\emm_v) \approx \rho_v \mathcal{B}(T1_v,T2_v)$. This network is trained separately from the encoder. Training uses physical (Bloch) simulations for many combinations of the T1, T2 and PD values  which can flexibly produce a rich training dataset~\cite{golbabaee2020compressive}.

\textbf{Encoder network:} projects  $\eg$ the gradient-updated TSMIs  in each iteration (i.e. the first line of \eqref{eq:PDG})  to the quantitative property maps $\emm$. Thus, $\G$ must simultaneously (i) learn to incorporate spatial-domain regularities to de-alias TSMIs from the undersampling artefacts, and (ii) resolve the temporal-domain \emph{inverse} mapping from the (noisy) TSMIs to the quantitative property maps. For this, and unlike \BLC\ which applies pixel-wise temporal-only processing, $\G$ uses multichannel convolution filters with wider receptive fields to learn/enable spatiotemporal processing of the TSMIs.

\begin{figure}[t]
\begin{center}
\includegraphics[width=1\linewidth]{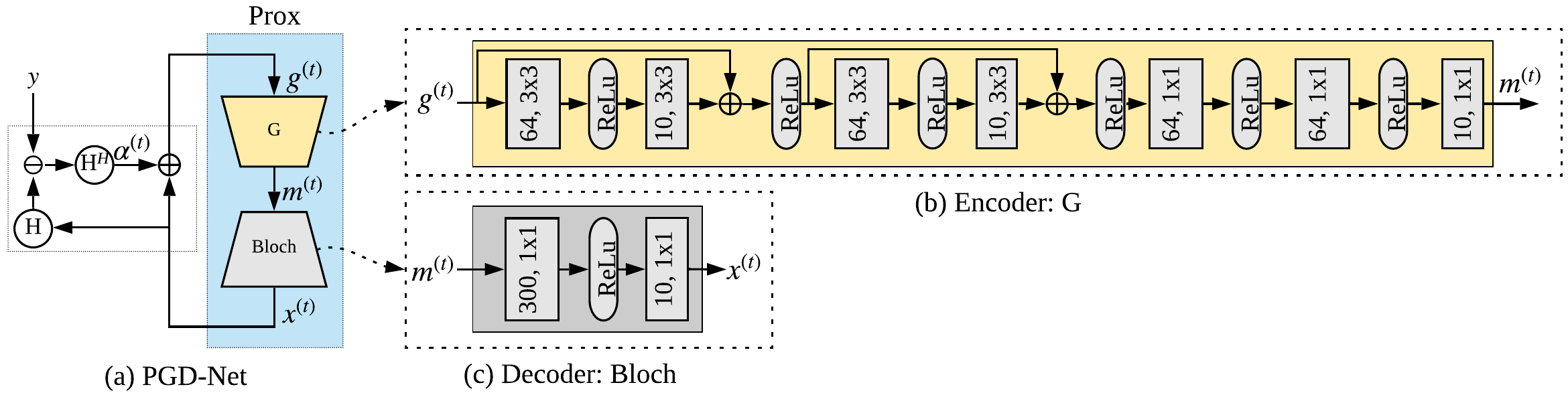}%0.48 0.82 0.06
\label{fig:compare_n}
\end{center}
\caption{Overview of the proposed proximal gradient descent network (PGD-Net) for tissue quantification in the compressive MR fingerprinting.}
\label{fig:RNN}
\end{figure}
\textbf{\proxnet:} Fig.~\ref{fig:RNN}a shows the \emph{recurrent} architecture of the proposed \emph{learned} PGD algorithm, coined as the \proxnet. The trainable parameters within the \proxnet\ are those of the encoder network $\G$ (Fig.~\ref{fig:RNN}b) and the step sizes $\alpha_t$. Other operators such as $\eH, \eH^H$ and \BLC\ (pre-trained separately, Fig.~\ref{fig:RNN}c) are kept frozen during training. Further, $\G$'s parameters are \emph{shared} through all iterations.  In practice, a truncated $T\geq1$ recurrent iterations is used for training. Supervised training requires the MRF measurements, TSMIs, and the ground truth property maps to form the training input $\ey$ and target $\ex, \emm$ samples.

Note there are many arts of engineering to determine the optimal network architecture, including different ways to encode temporal \cite{hoppe2017deep} or spatial-temporal information \cite{balsiger2019spatial}, these aspects are somewhat orthogonal to the model consistency question. Indeed, such mechanisms could also be incorporated in \proxnet.

\textbf{Training loss:}
Given a training set $\{\ex_i, \emm_i, \ey_i\}_{i=1}^N$, and $T\geq 1$ recurrent iterations of the \proxnet\  (i.e. iterations used in PGD), the loss is defined as
\begin{equation}\label{eqs:loss}
\mathcal{L} =  \gamma\sum\limits_{i=1}^N\ell\left(\ex_i, \ex_i^{(T)}\right) +  \sum\limits_{j\in\{T_1, T_2, \rho \}}\beta_j \sum\limits_{i=1}^N \ell\left(\emm_{ij},  \emm^{(T)}_{ij}\right) + \lambda \sum\limits_{t=1}^T \sum\limits_{i=1}^N\ell\left(\ey_i, \eH(\ex_i^{(t)})\right),
\end{equation}
where $\ell$ is the MSE loss defined with appropriate weights $\gamma$, $\beta_j, \lambda$ on the reconstructed TSMIs $\ex$ (which measures the Bloch dynamic consistency) and tissue property maps $\emm$, as well as on $\ey$ to maximise k-space data consistency with respect to the (physical) forward acquisition model. In this paper, the scaling between parameters $\gamma$, $\beta_j$ and $\lambda$ were initialized based on the physics (see \ref{sec:rec_algos}).

\section{Numerical experiments}
\subsection{Anatomical dataset}
We construct a dataset of brain scans acquired using the 1.5T GE HDxT scanner with 8-channel receive-only head RF coil. For setting ground-truth (GT) values for the T1, T2 and PD parameters, gold standard anatomical maps were acquired using MAGIC quantification protocol~\cite{magic2015}. Ground-truth quantitative maps were acquired from 8 healthy volunteers (16 axial brain slices each, at the spatial resolution of $128\times 128$ pixels). From these parametric maps, we then construct the TSMIs and MRF measurements using the MRF acquisition protocol mentioned below to form the training/testing tuples $(\emm_i,\ex_i,\ey_i)$. Data from 7 subjects were used for training our models, and one subject was kept for performance testing. We augmented training data into total 224 samples using random rotations (uniform angles in $[-8^\circ,8^\circ]$), and left-right flipping of the GT maps. Training batches at each learning epoch were corrupted by i.i.d Gaussian noises of 30dB SNR added to $\ey$ (we similarly add noise to the k-space test data).

\subsection{MRF acquisition}
Our experiments use an excitation sequence of $L=200$ repetitions which jointly encodes T1 and T2 values using an inversion pulse followed by a flip angle schedule that linearly ramps up from $1^\circ$ to $40^\circ$, i.e. $\times4$ truncated sequence than~\cite{gomez2020rapid, golbabaee2020compressive}. Following~\cite{gomez2020rapid}, we set acquisition parameters  Tinv=18 msec (inversion time), fixed TR=10 msec (repetition time), and TE = 0.46 msec (echo time).  Spiral readouts subsample the k-space frequencies (the $128\times128$ Cartesian FFT grid) across 200 repetition times.
We  sample spatial frequencies $k(t)\propto(1.05)^{\frac{16\pi \tau}{1000}}e^{j\frac{16\pi \tau}{1000}}$ for $\tau=1,2,...,1000$, which after quantisation to the nearest FFT grid, results in $m=654$ samples per timeframe. In every repetition, similar to \cite{ma2013magnetic}, this spiral pattern rotates by $7.5^\circ$ in order to sub-sample new k-space frequencies. Given the anatomical T1, T2 and PD maps, we simulate magnetic responses using the Extended Phase Graph (EPG) formalism \cite{weigel2015extended} and construct TMSIs and k-space measurements datasets, and use them for training and retrospective validations.

\subsection{Reconstruction algorithms}\label{sec:rec_algos}
Two DM baselines namely, the non-iterative Fast Group Matching (FGM) \cite{cauley2015fast} and the model-based iterative algorithm BLIP empowered by the FGM's fast searches, were used for comparisons. For this, a MRF dictionary of 113'781 fingerprints was simulated over a dense grid of (T1, T2)=[100:10:4000]$\times$[20:2:600]~msec values. We implemented FGM searches on GPU using 100 groups for clustering this dictionary. The BLIP algorithm uses backtracking step size search and runs for maximum 20 iterations if is not convergent earlier. Further, we compared against related deep learning MRF baselines MRFCNN \cite{chen2019deep} and SCQ \cite{fang2019deep}. In particular, MRFCNN is a fully convolutional network and SCQ mainly uses 3  U-nets to separately infer T1, T2 and PD maps.
 The input to these networks is the dimension-reduced back-projected TSMIs $\eH^{H}(\ey)$, and their training losses only consider quantitative maps consistency i.e. the second term in~\eqref{eqs:loss}.

We trained \proxnet\ with recurrent iterations $T=2$ and 5 to learn appropriate proximal encoder $\G$ and the step sizes $\alpha^{(t)}$.
The architectures of $\G$ and \BLC\ networks are illustrated in Fig.~\ref{fig:RNN}. Similar to~\cite{golbabaee2020compressive}, the MRF dictionary was used for pre-training the \BLC\ decoder that embeds a differentiable model for generating Bloch magnetic responses. A compact shallow network with one hidden layer and $1\times1$ filters (for pixel-wise processing) implements our \BLC\ model~\cite{golbabaee2020compressive}. On the other hand, our encoder $\G$ has two residual blocks with $3\times 3$ filters (for de-aliasing) followed by three convolutional layers with $1\times 1$ filters for quantitative inference. The final hyper-parameters were $\beta=[1, 20, 2.5]$, $\gamma=10^{-3}$ and $\lambda=10^{-2}$ selected via a multiscale grid search to minimize error w.r.t. the ground truth. The inputs were normalized such that PD ranged in $[0, 1]$; smaller weights were used for $\ex$ and $\ey$ since they have higher energy than PD; we set $\lambda > \gamma$ since $\ex$’s norm is larger than $\ey$; $T1/T2$ values typically exhibit different ranges with $T1\gg T2$, justifying their relative weightings in $\beta$ to balance these terms. We used ADAM optimiser with 2000 epochs, mini-batch size 4 and learning rate $10^{-4}$. We pre-trained our encoder $\G$ using back-projected TSMIs to initialise the recurrent training, and also to compare the \textit{encoder alone} predictions  to the \proxnet. All algorithms use a $s=10$ dimensional MRF subspace representation for temporal-domain dimensionality reduction. The input and output channels are respectively 10 and 3 for MRFCNN, SCQ and $\G$.  All networks  were implemented in PyTorch, and trained and tested on NVIDIA 2080Ti GPUs.

\begin{table}[t!]
	\centering
	\fontsize{8}{8}\selectfont
		\caption{ Average errors (NRMSE, SSIM, MAE), memory (for storing a dictionary or a network) and runtimes (per image sclice) required for computing T1, T2 and PD maps using the MRF baselines and our \proxnet\ algorithm.
}
	\scalebox{.93}{
		\begin{tabular}{lccc|ccc|cc|c|c}%{cp{.8cm}p{.8cm}p{.9cm}p{.9cm}}
			\toprule[0.2em]
			&\multicolumn{3}{c}{NRMSE} & \multicolumn{3}{c}{SSIM} & \multicolumn{2}{c}{MAE (msec)} &  \multicolumn{1}{c}{time (sec)} & \multicolumn{1}{c}{memory (MB)} \\
			\midrule[0.1em]
			&  {T1 } & { T2 } & PD &    {T1 } & { T2 } &  PD& {T1 } & { T2 }  \\
			\midrule[0.05em]
			\midrule[0.05em]
			FGM &  0.475 & 0.354 & 1.12 & 0.614 &0.652 & 0.687 & 350.0 & 14.6 &1.29 & 8.81 \\
			\midrule[0.05em]
			BLIP+FGM & 0.230 & 0.545 & 0.073 & 0.886 & 0.880 & 0.984 & 91.7 & 8.0 &79.28 & 8.81 \\	
			\midrule[0.05em]
			MRFCNN & 0.155 & 0.158 & 0.063 & 0.943 & 0.972 & 0.987 & 80.3 & 5.4 & 0.083 & 4.72 \\	
			\midrule[0.05em]
			SCQ & 0.172 & 0.177& 0.064 & 0.929 & 0.967 & 0.984 & 91.7 & 6.1 & 0.132 & 464.51\\								
			\midrule[0.05em]
			$\G$ (encoder alone)  & 0.142 & 0.155 & 0.065  & 0.948 & 0.973 & 0.987 & 77.1 & 5.6 & \textbf{0.067} & \textbf{0.55} \\														
			\midrule[0.05em]
			\proxnet\ ($T=2$)   & 0.104 & 0.138 & 0.050  & 0.973 & 0.979 &0.991 & 59.9 & 5.0& 0.078& 0.57\\	
			\midrule[0.05em]
			\proxnet\ ($T=5$)  & \textbf{0.100} & \textbf{0.132} & \textbf{0.045} & \textbf{0.975} & \textbf{0.981} & \textbf{0.992} & \textbf{50.8} & \textbf{4.6} & 0.103 &0.57\\	
			\bottomrule[0.2em]
		\end{tabular}}
 \label{tab:testdata1}
	\end{table}

\begin{figure*}[ht!]
	\centering
	\scalebox{1}{
	\begin{minipage}{\linewidth}
		\centering		
			\includegraphics[width=.162\linewidth]{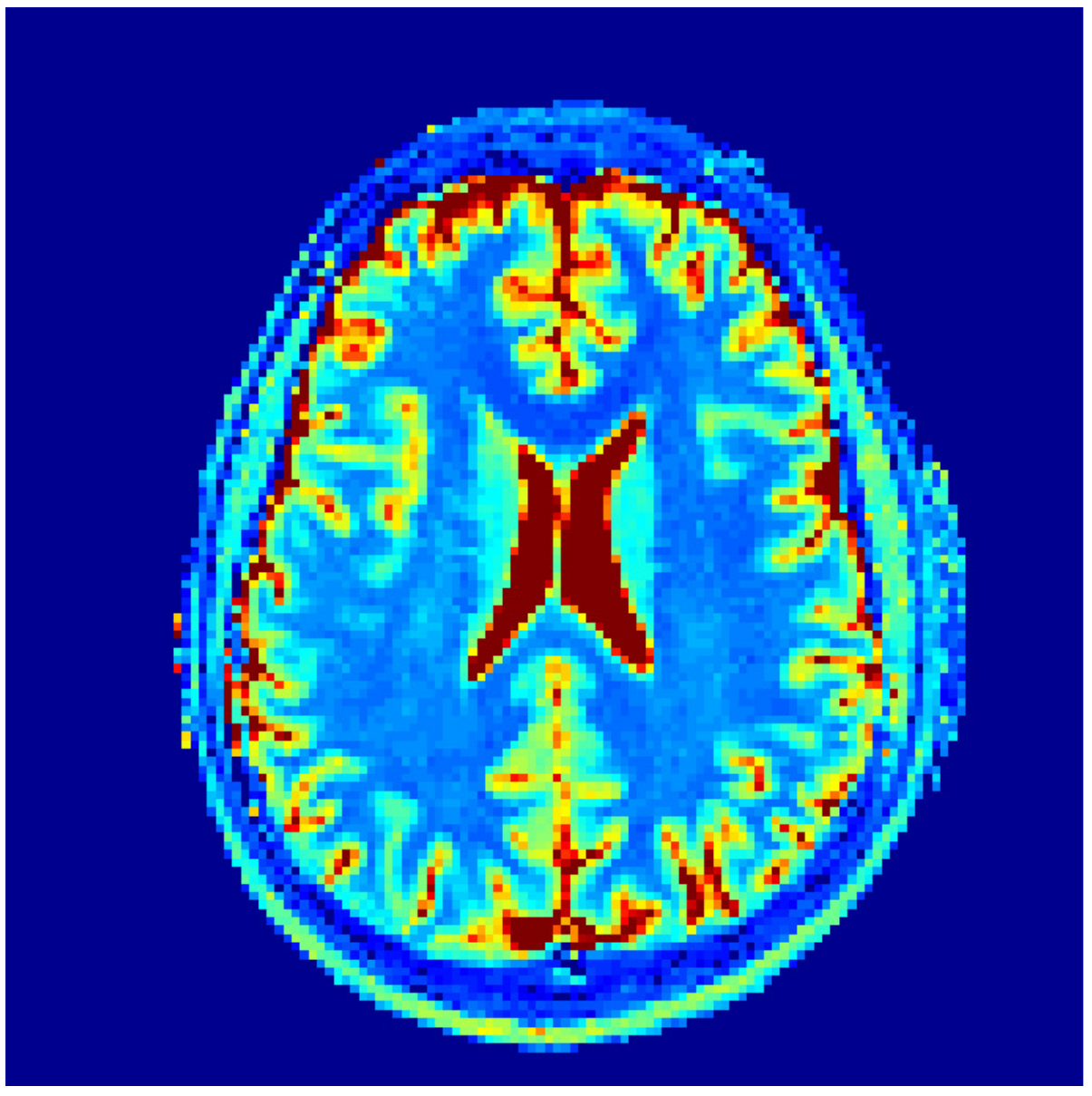}\hspace{-.1cm}
	\includegraphics[width=.162\linewidth]{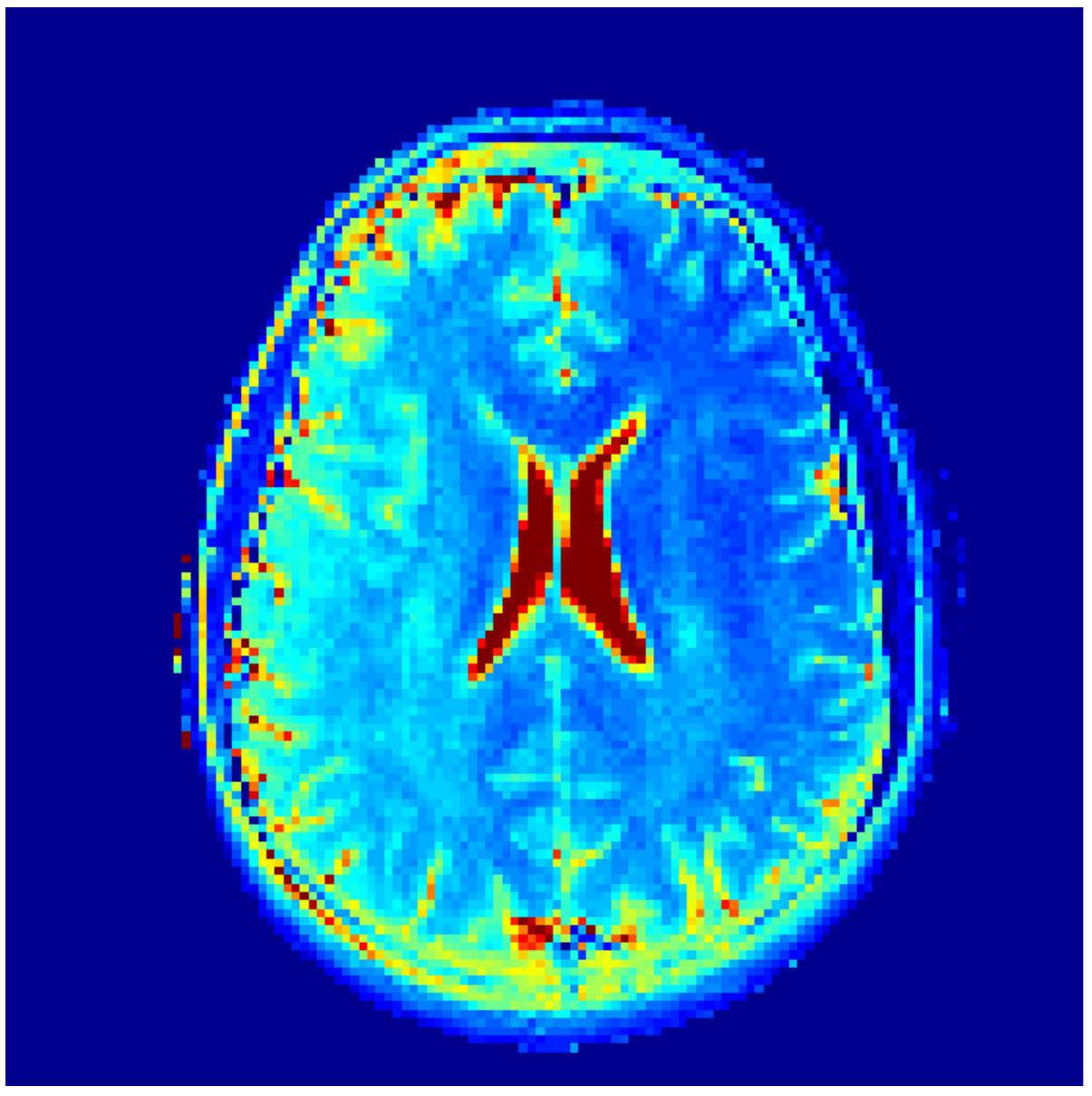}\hspace{-.1cm}
	\includegraphics[width=.162\linewidth]{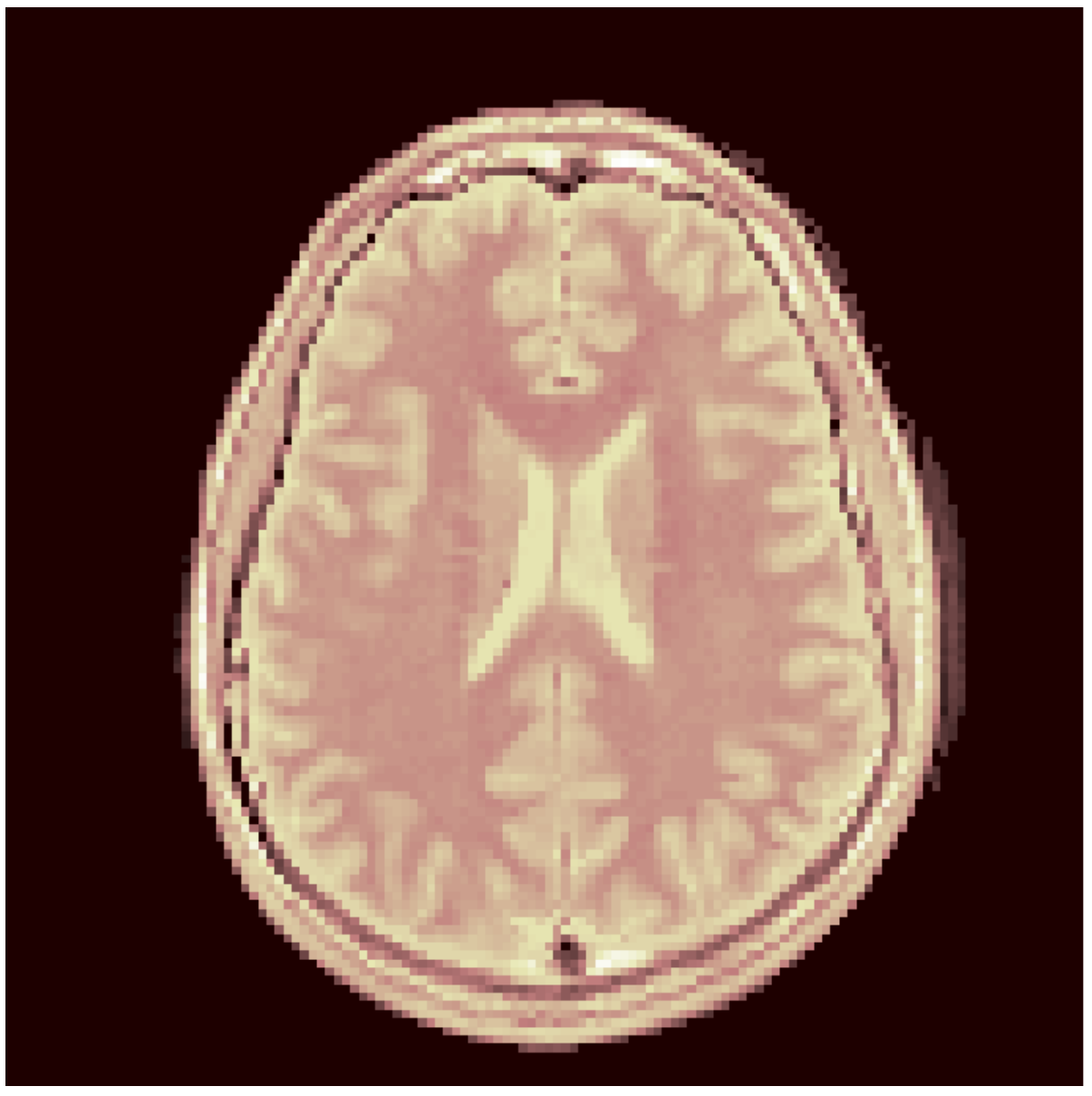}\hspace{-.1cm}
\\
\footnotesize{Ground truth (T1, T2, PD) anatomical maps acquired by the MAGIC gold standard~\cite{magic2015} }
\\
		\includegraphics[width=.162\linewidth]{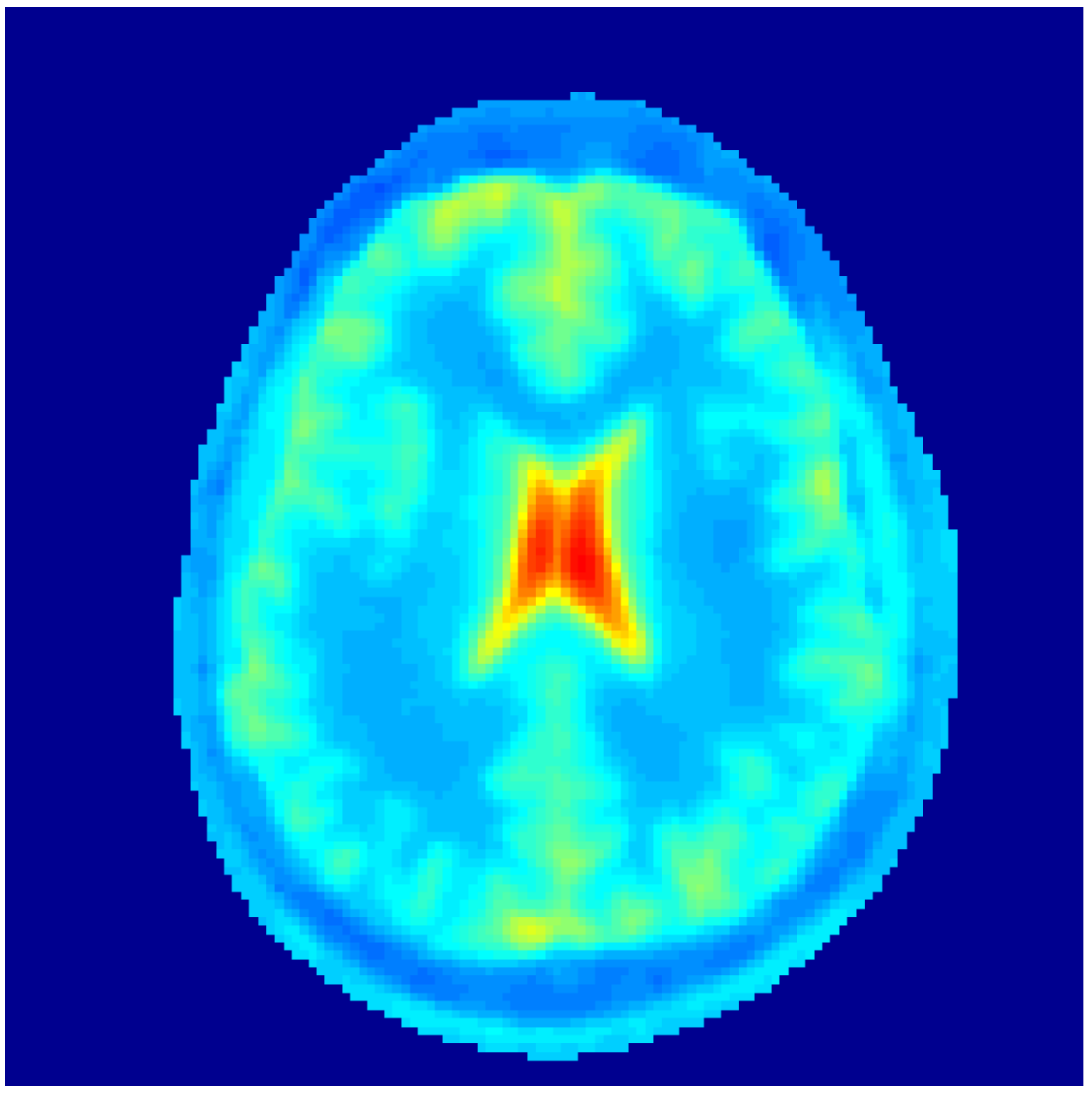}\hspace{-.1cm}
		\includegraphics[width=.162\linewidth]{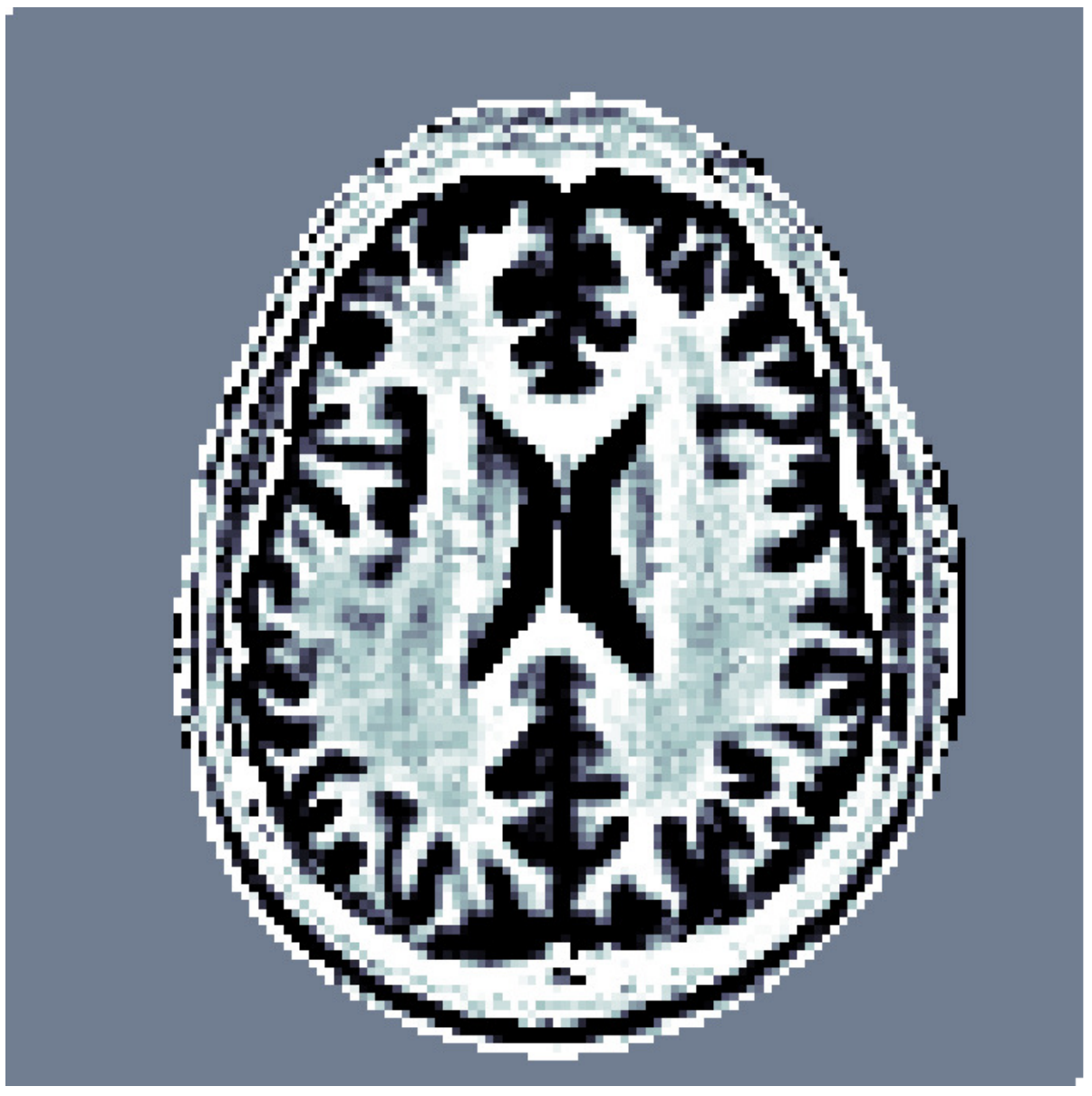}\hspace{.0cm}
		\includegraphics[width=.162\linewidth]{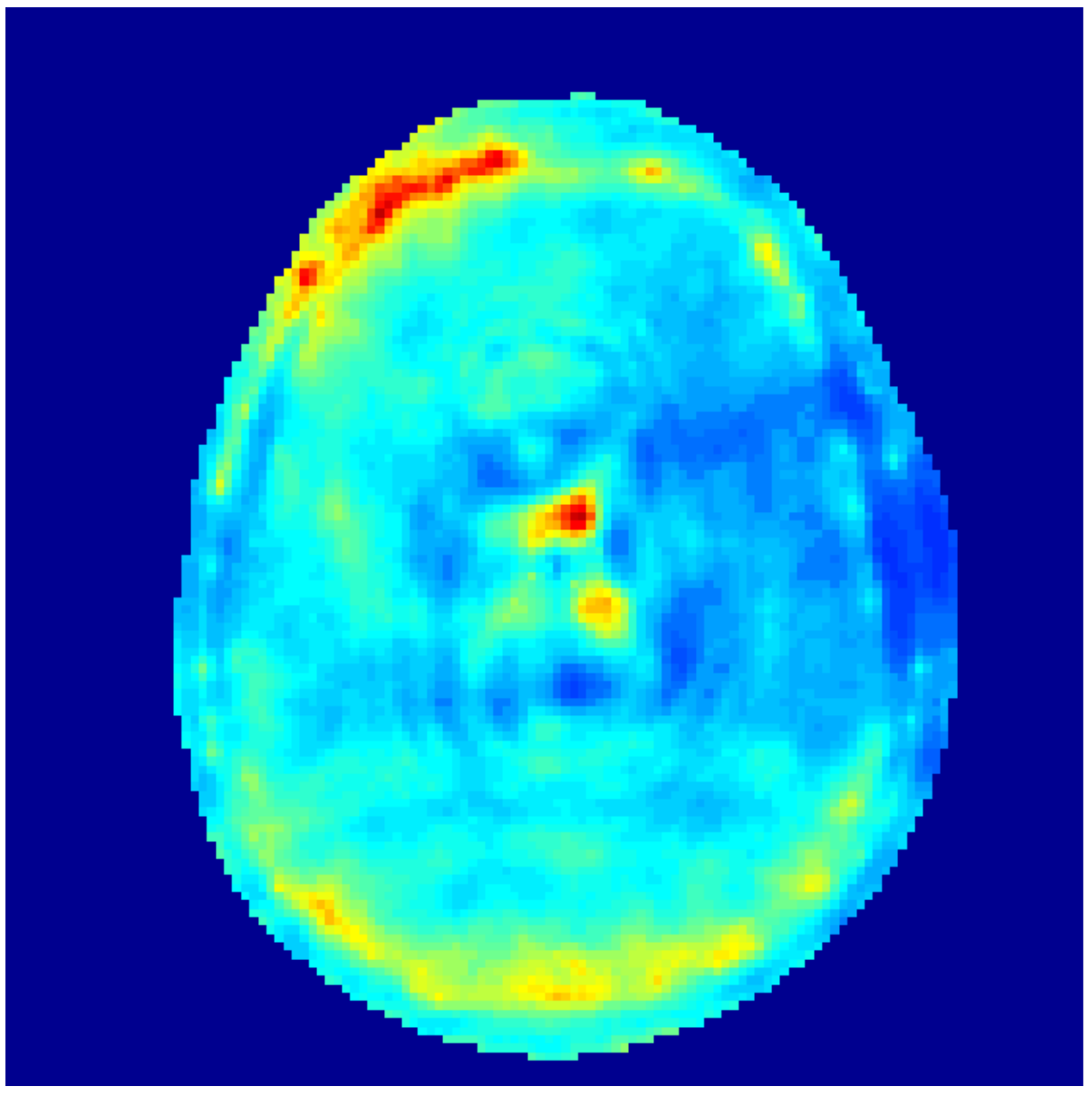}\hspace{-.1cm}
		\includegraphics[width=.162\linewidth]{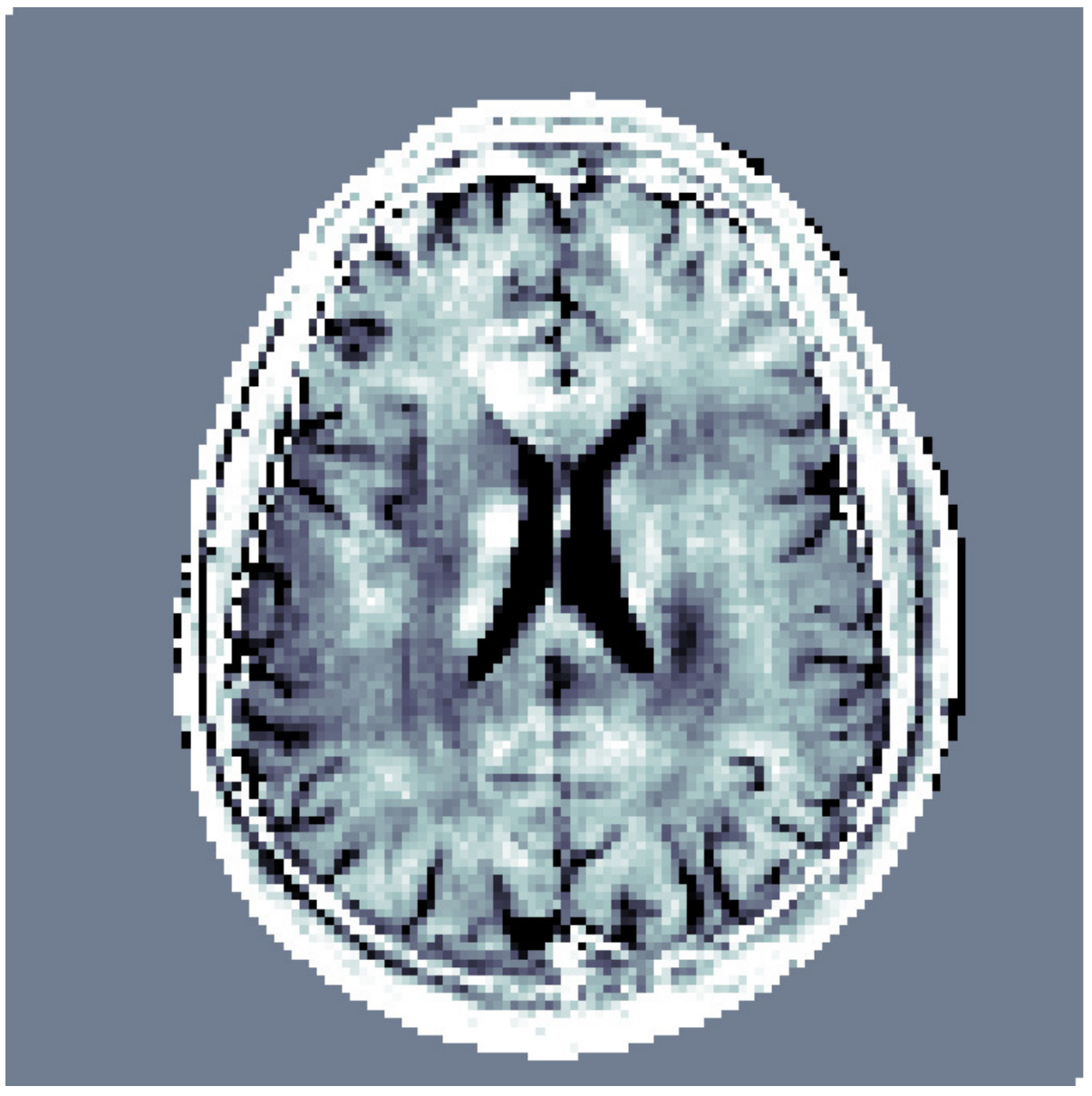}\hspace{.0cm}
		\includegraphics[width=.162\linewidth]{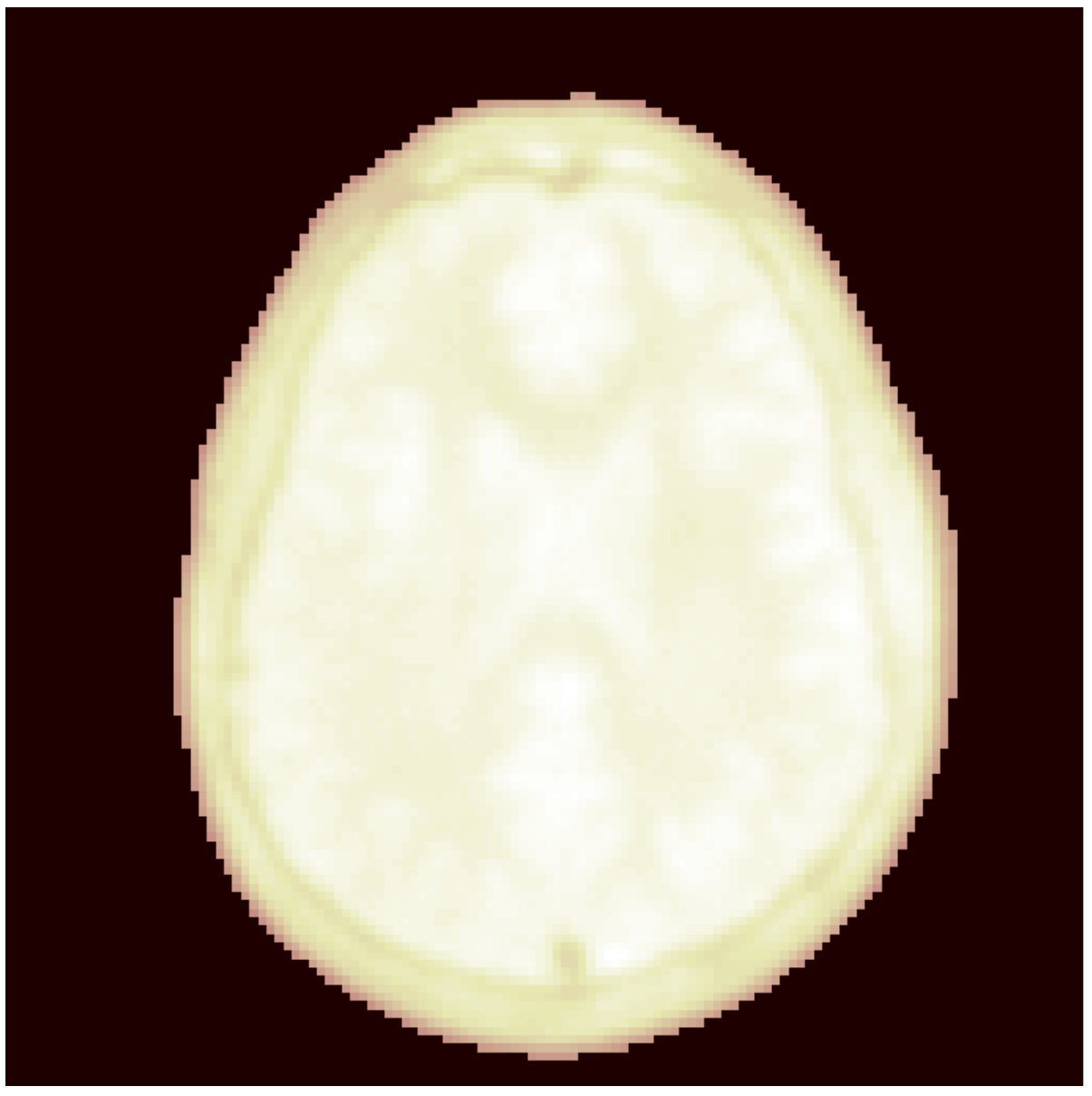}\hspace{-.1cm}
		\includegraphics[width=.162\linewidth]{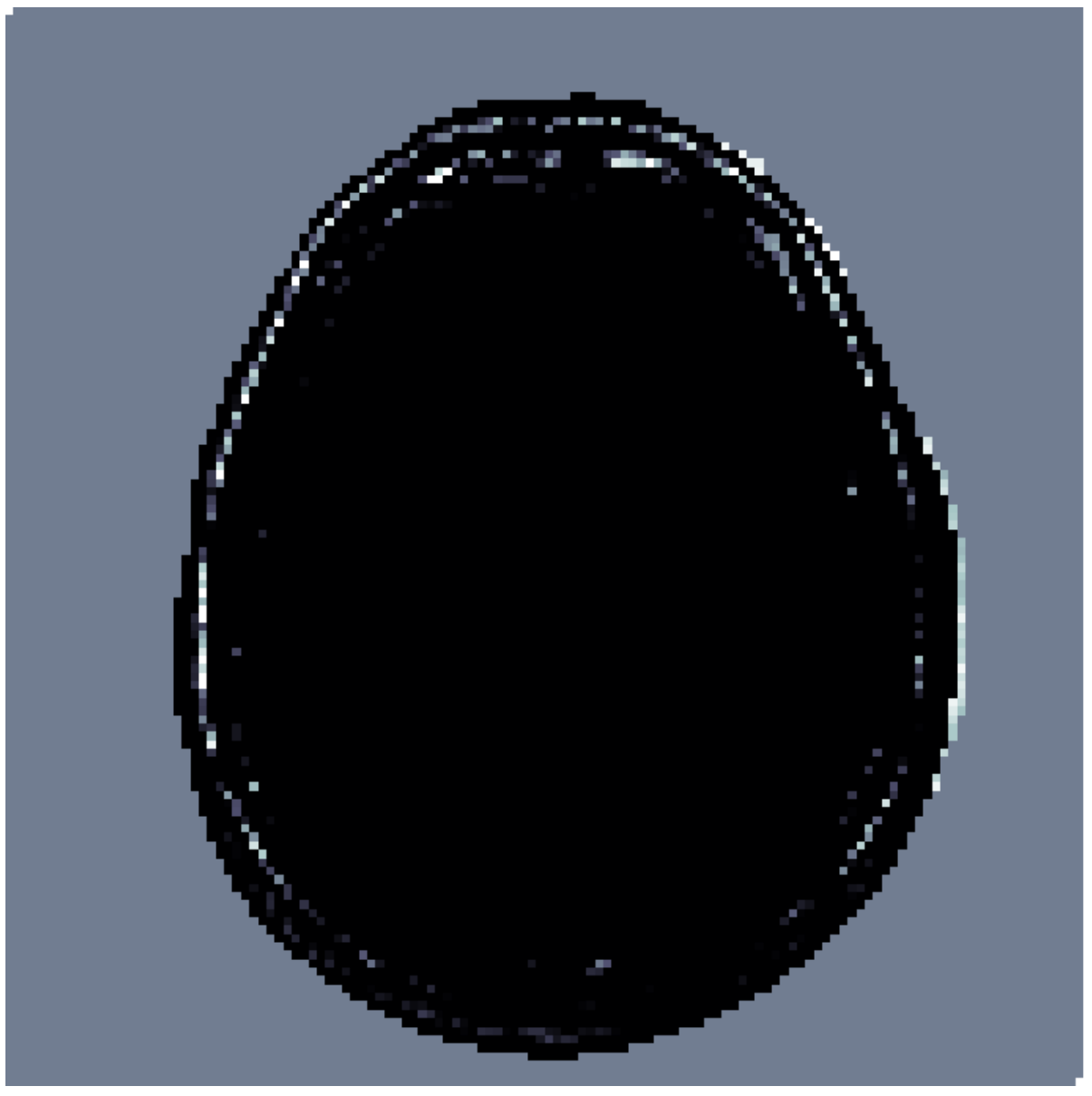}		
		\\
		\includegraphics[width=.162\linewidth]{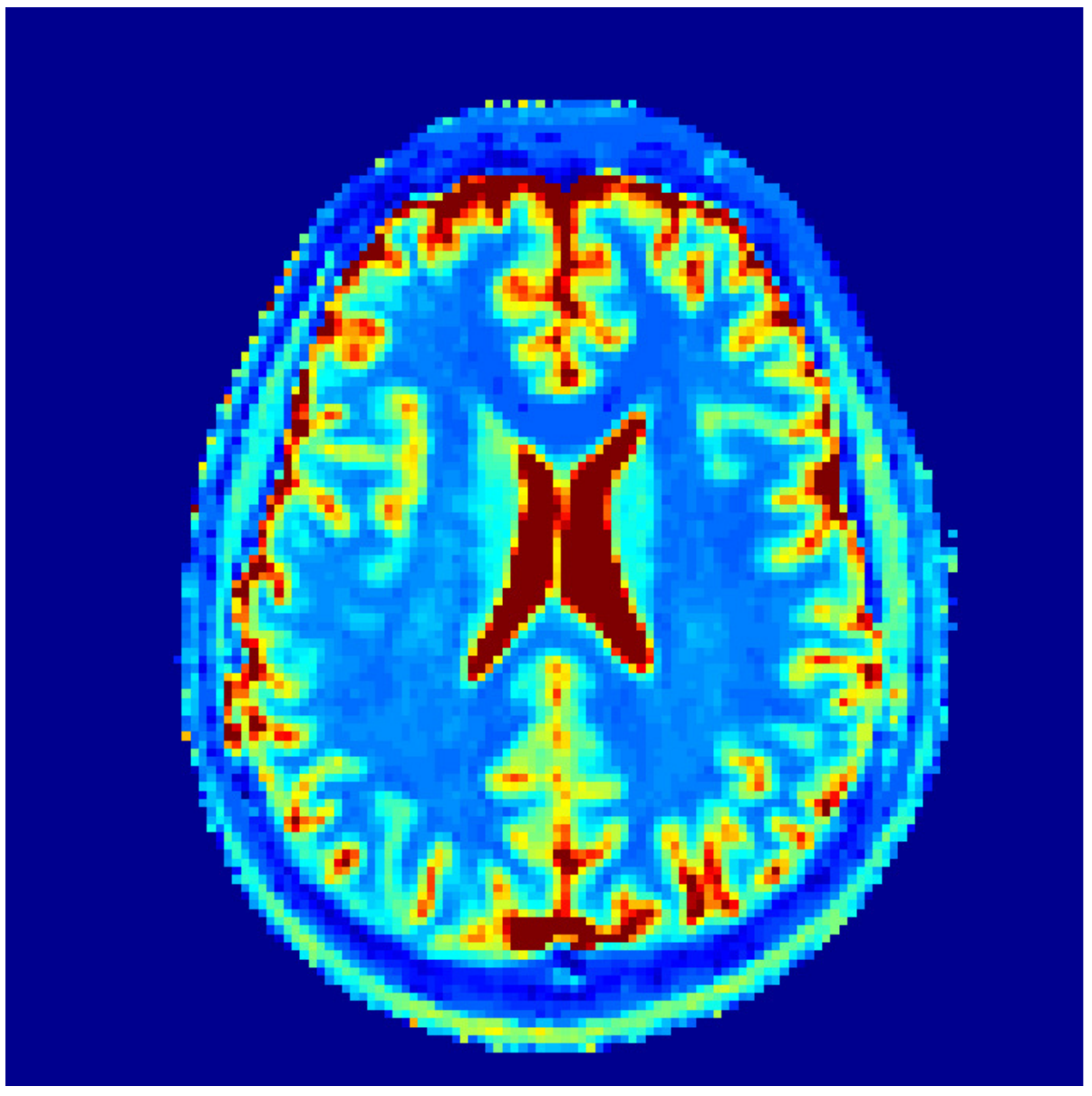}\hspace{-.1cm}
		\includegraphics[width=.162\linewidth]{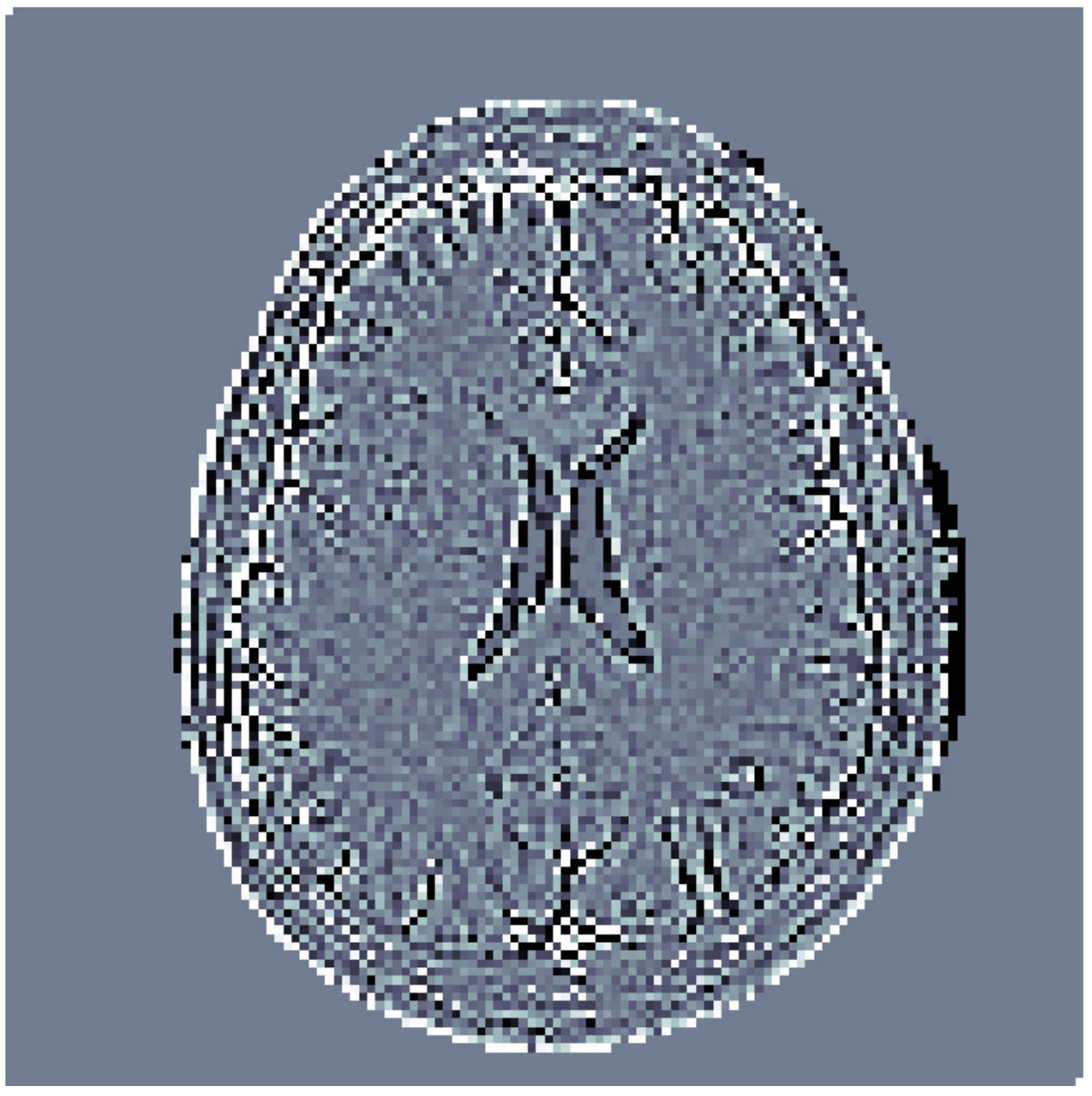}\hspace{.0cm}
		\includegraphics[width=.162\linewidth]{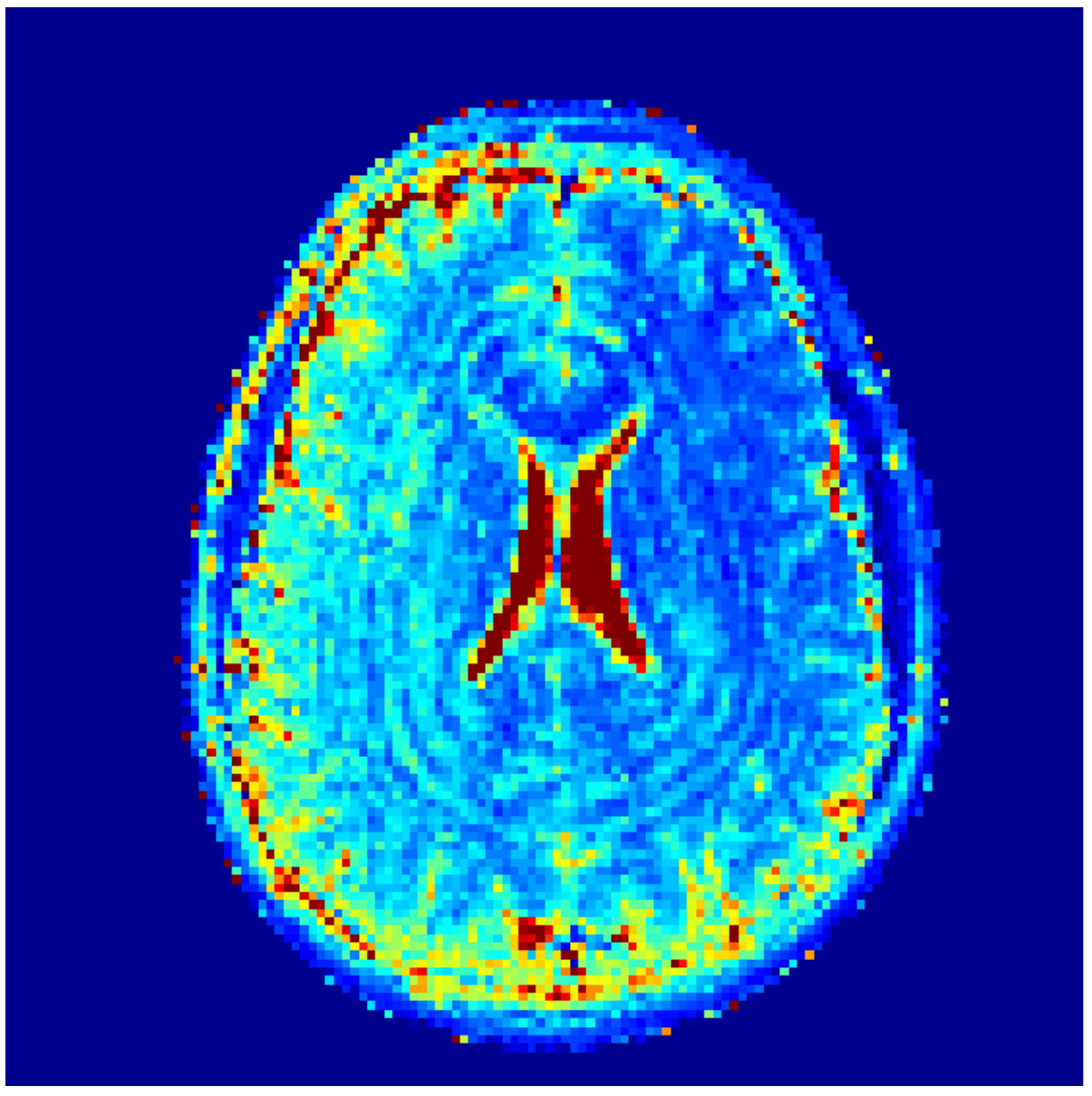}\hspace{-.1cm}
		\includegraphics[width=.162\linewidth]{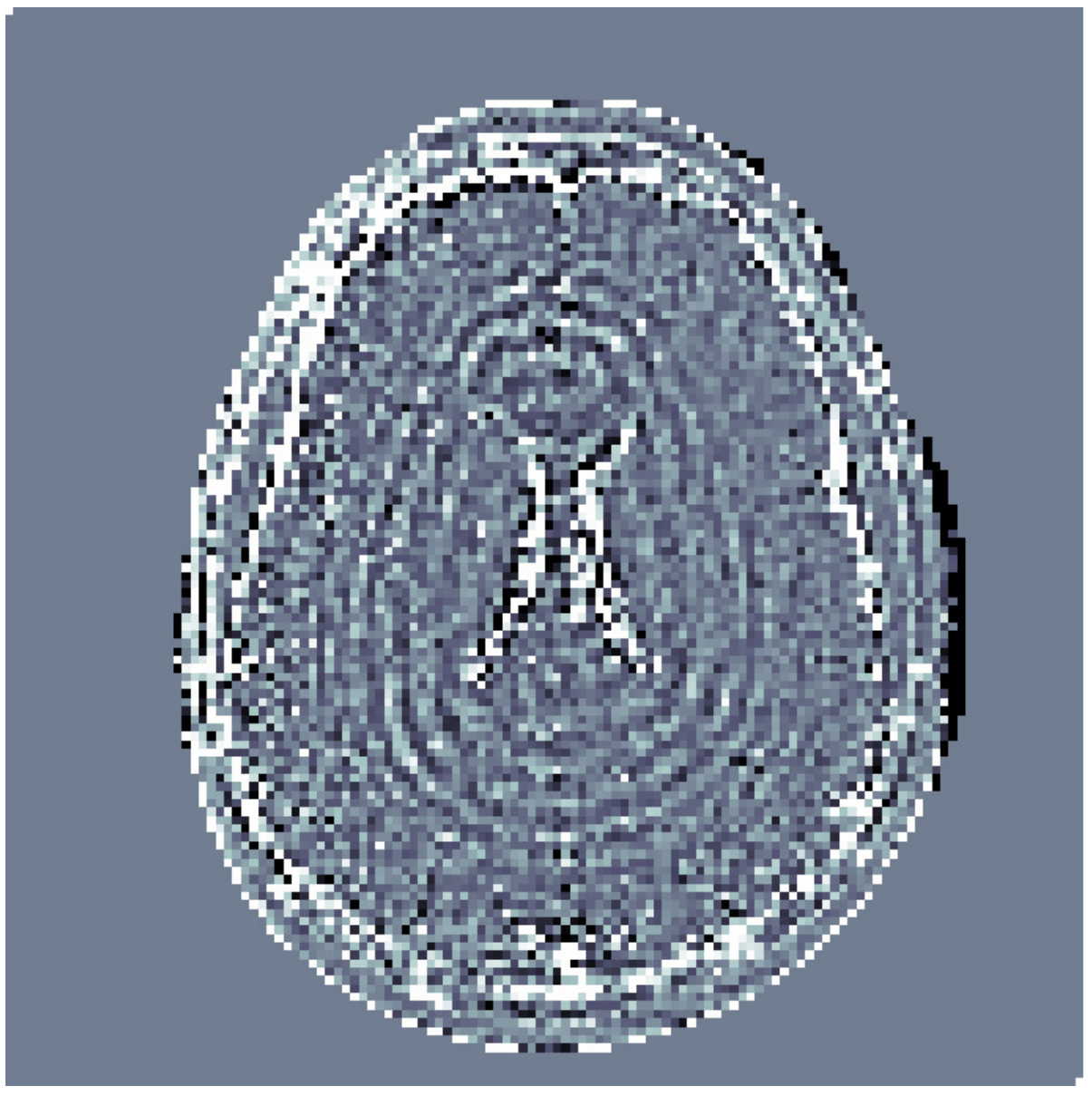}\hspace{.0cm}
		\includegraphics[width=.162\linewidth]{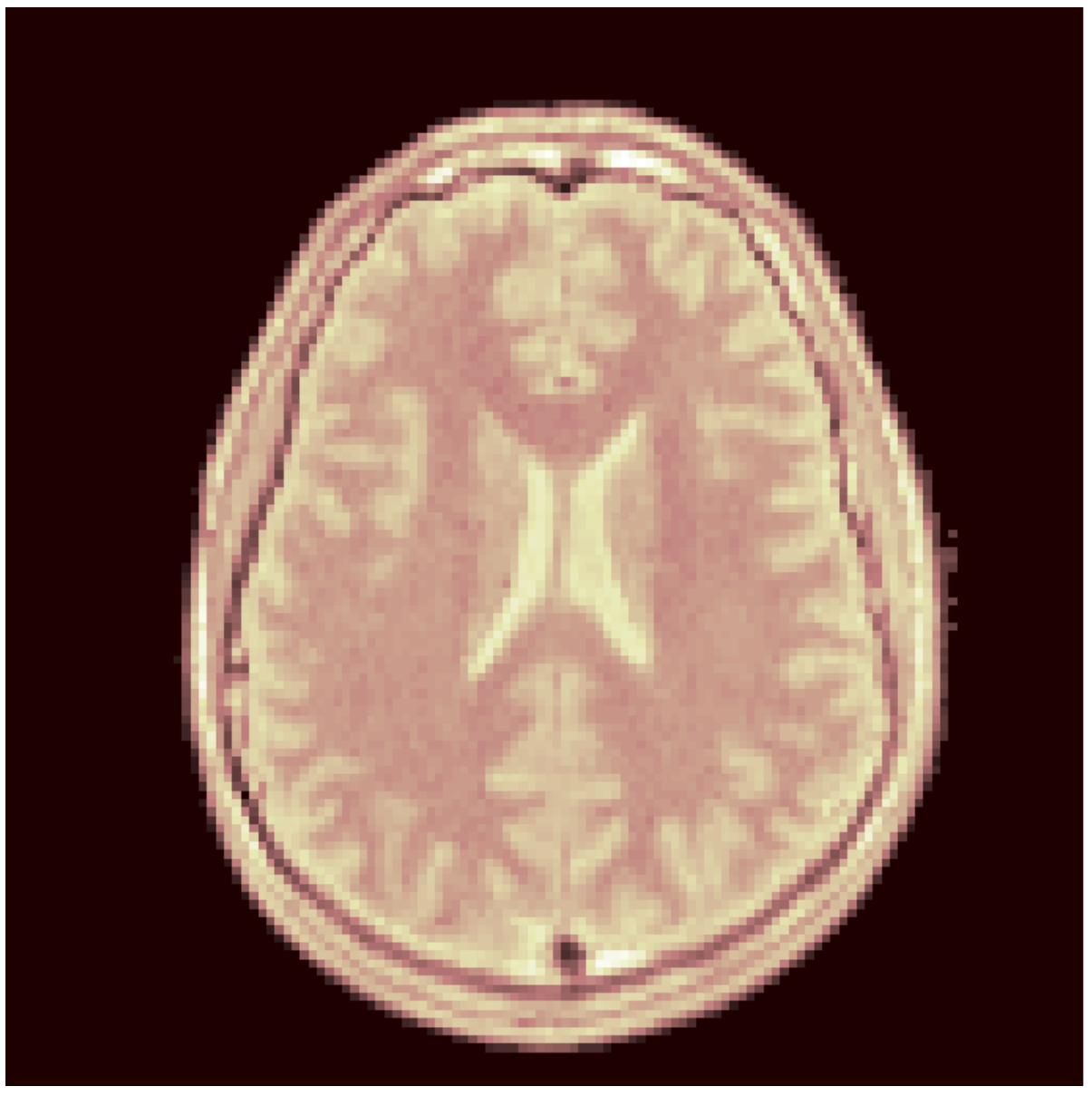}\hspace{-.1cm}
		\includegraphics[width=.162\linewidth]{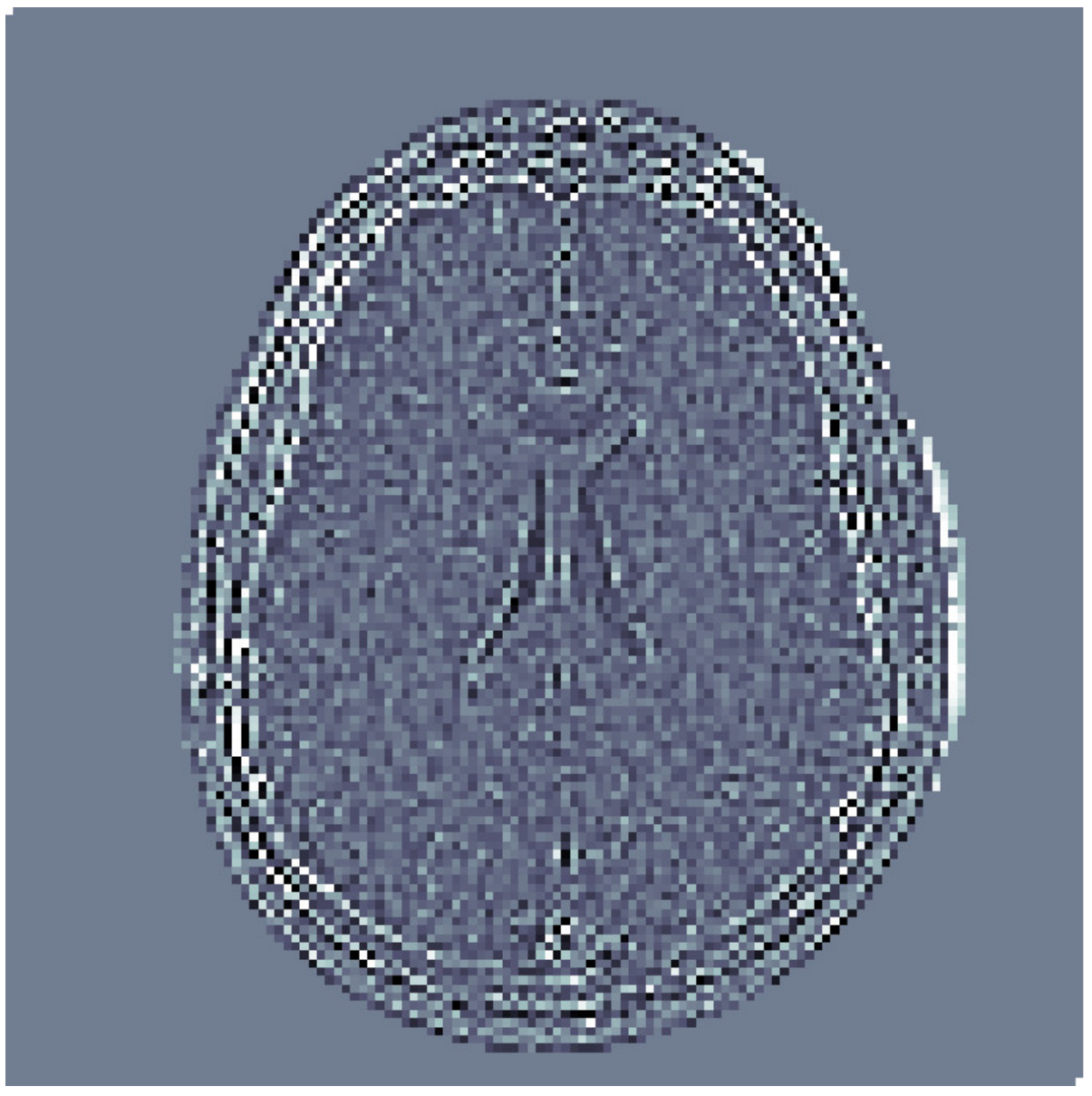}			
				\\
		\includegraphics[width=.162\linewidth]{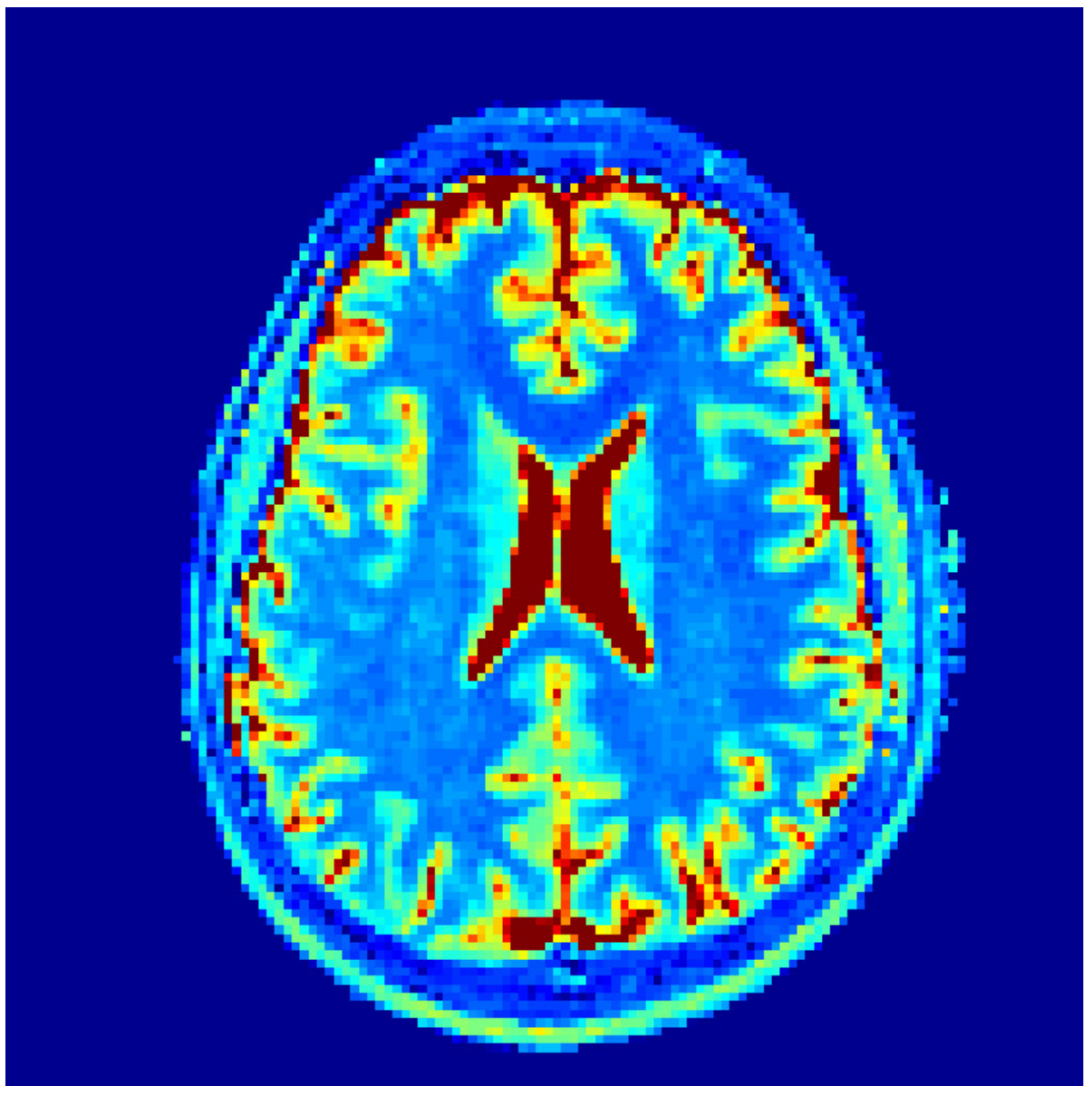}\hspace{-.1cm}
		\includegraphics[width=.162\linewidth]{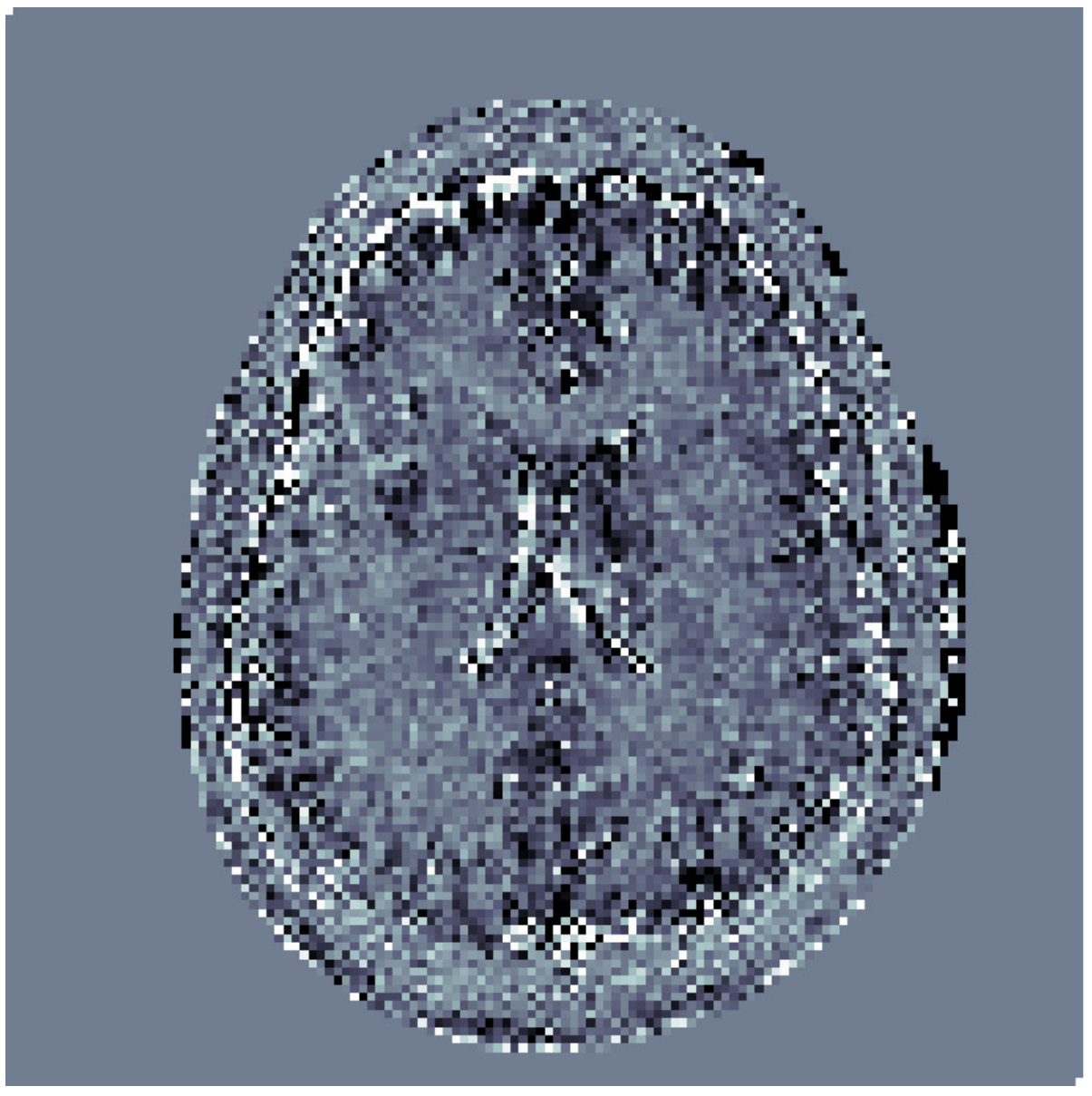}\hspace{.0cm}
		\includegraphics[width=.162\linewidth]{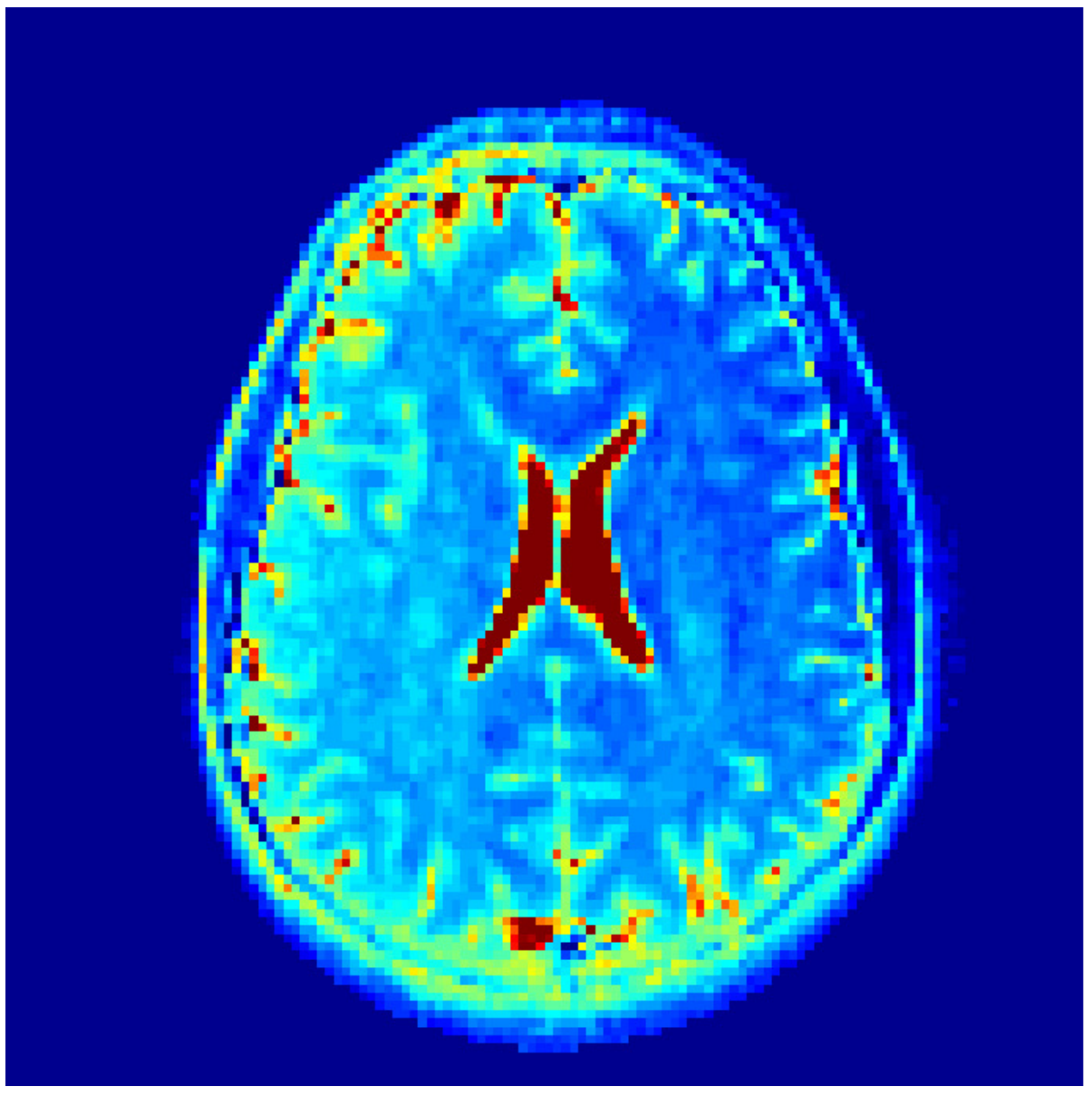}\hspace{-.1cm}
		\includegraphics[width=.162\linewidth]{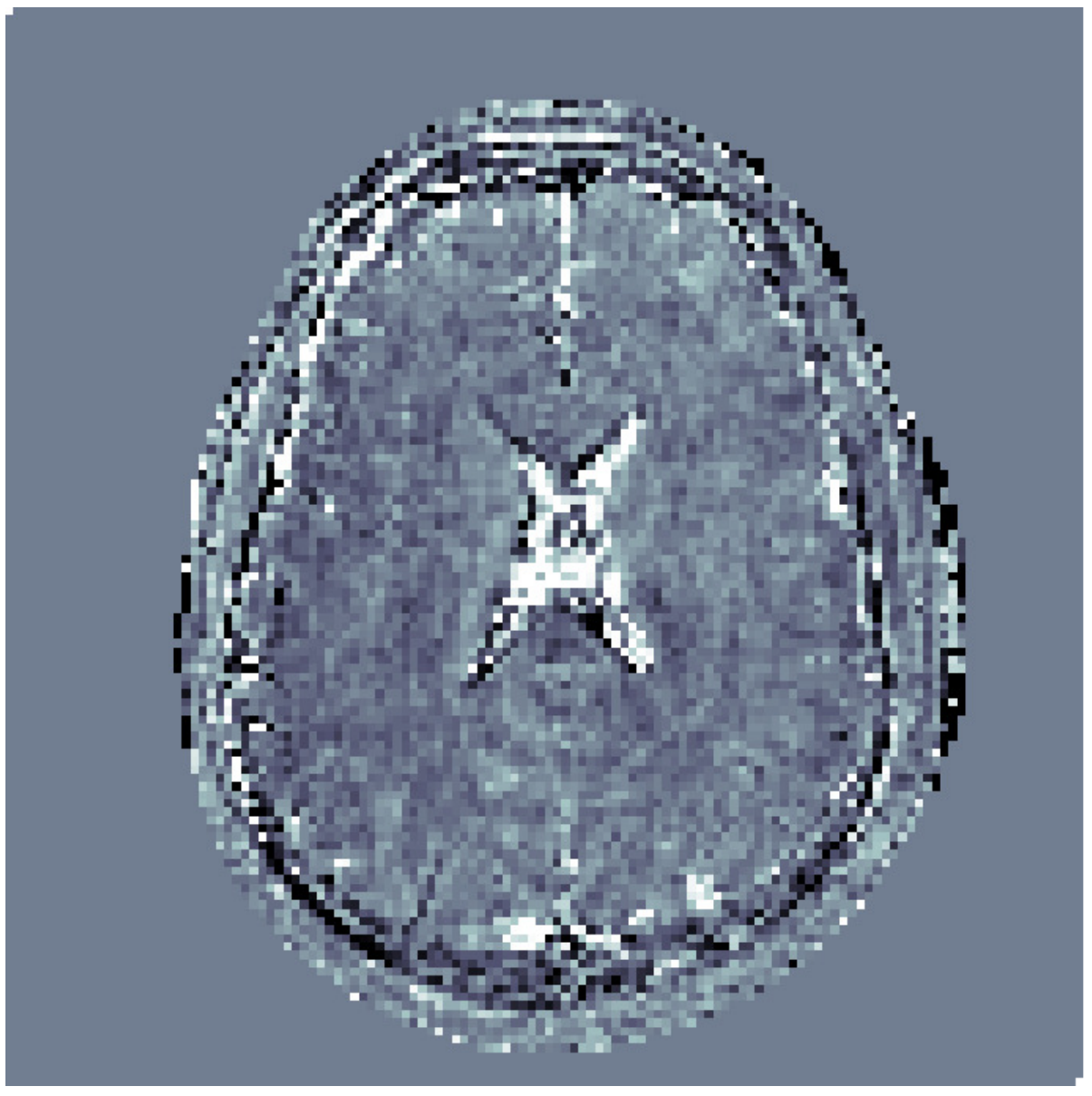}\hspace{.0cm}
		\includegraphics[width=.162\linewidth]{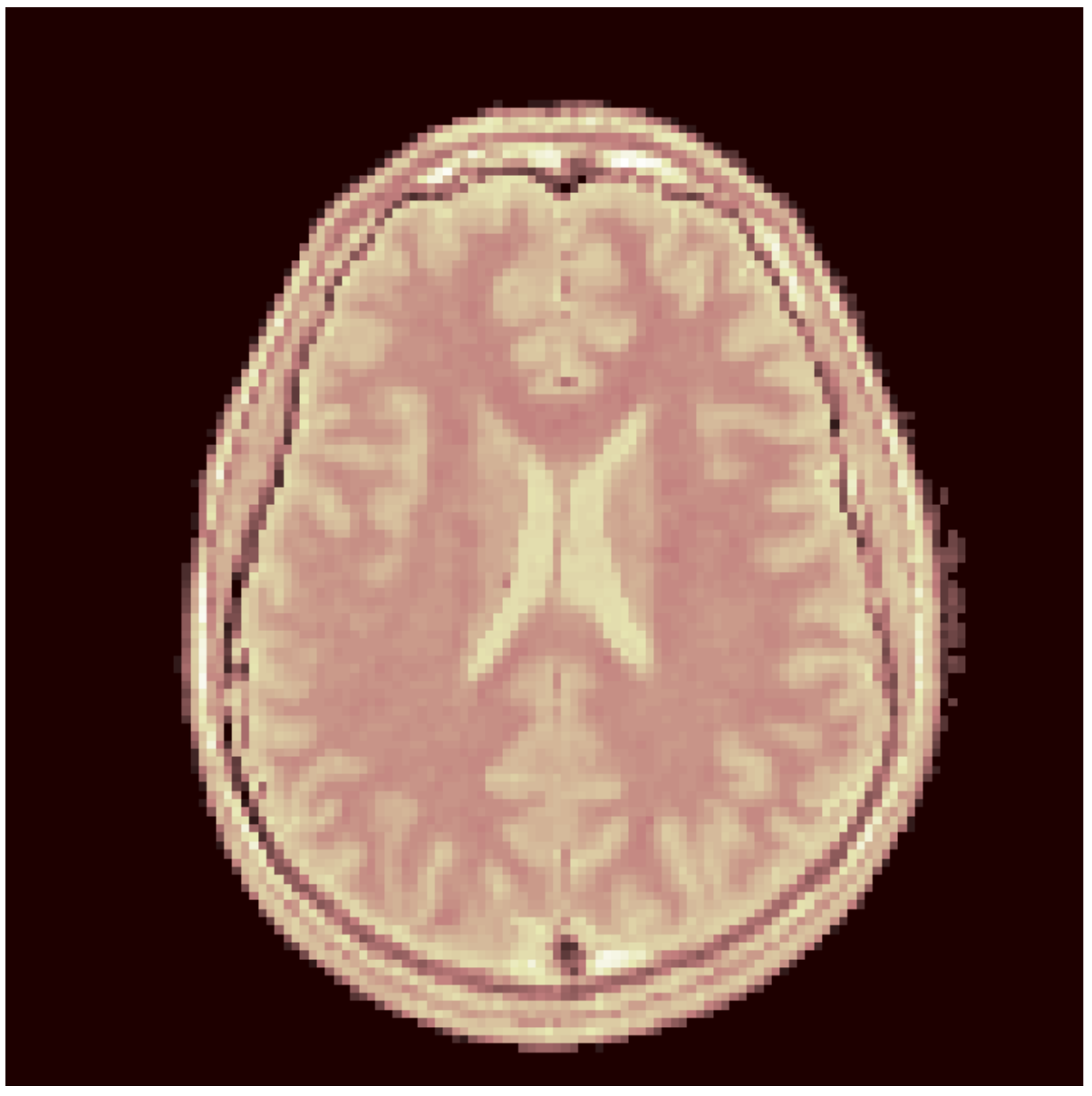}\hspace{-.1cm}
		\includegraphics[width=.162\linewidth]{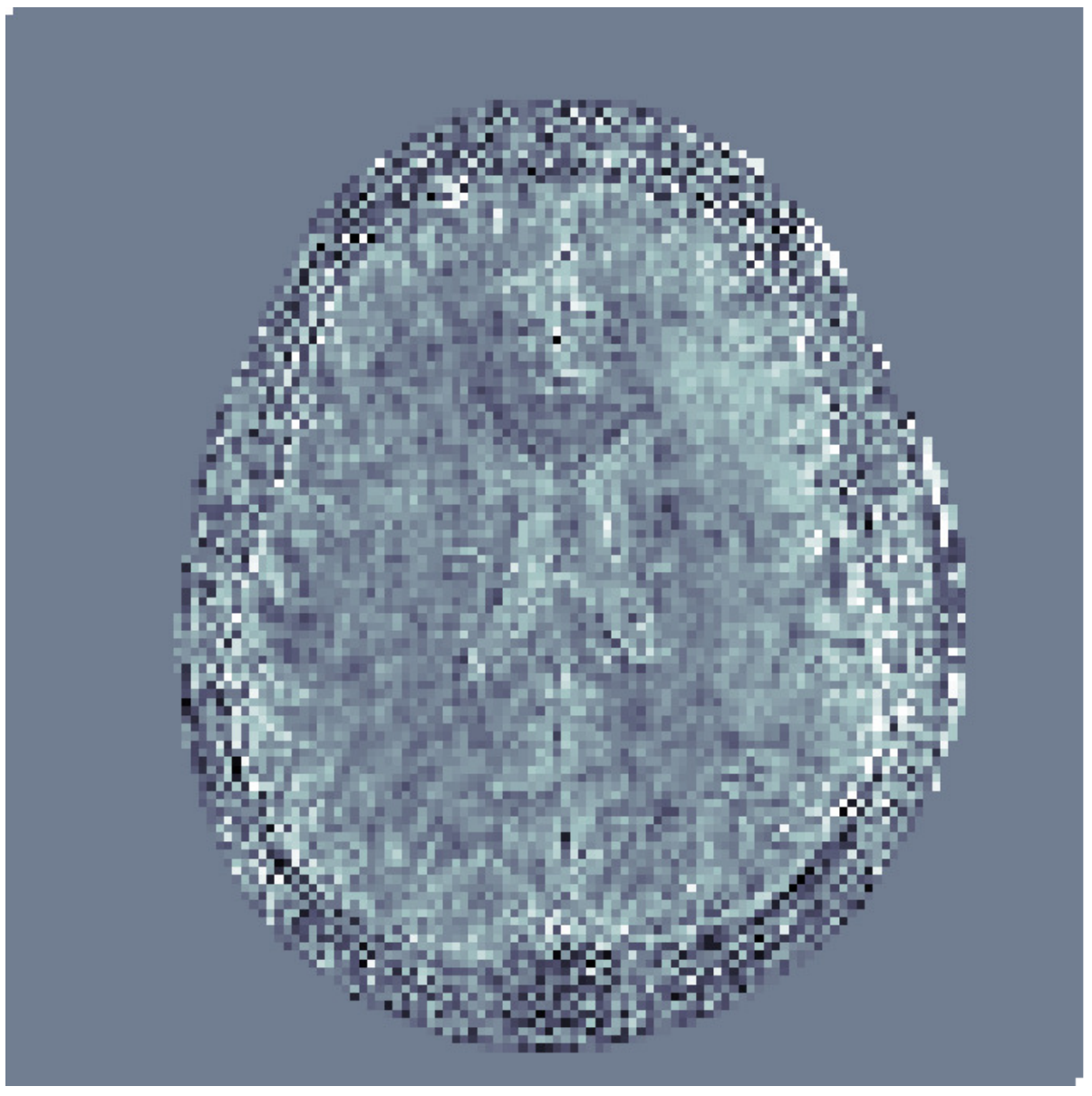}			
				\\
		\includegraphics[width=.162\linewidth]{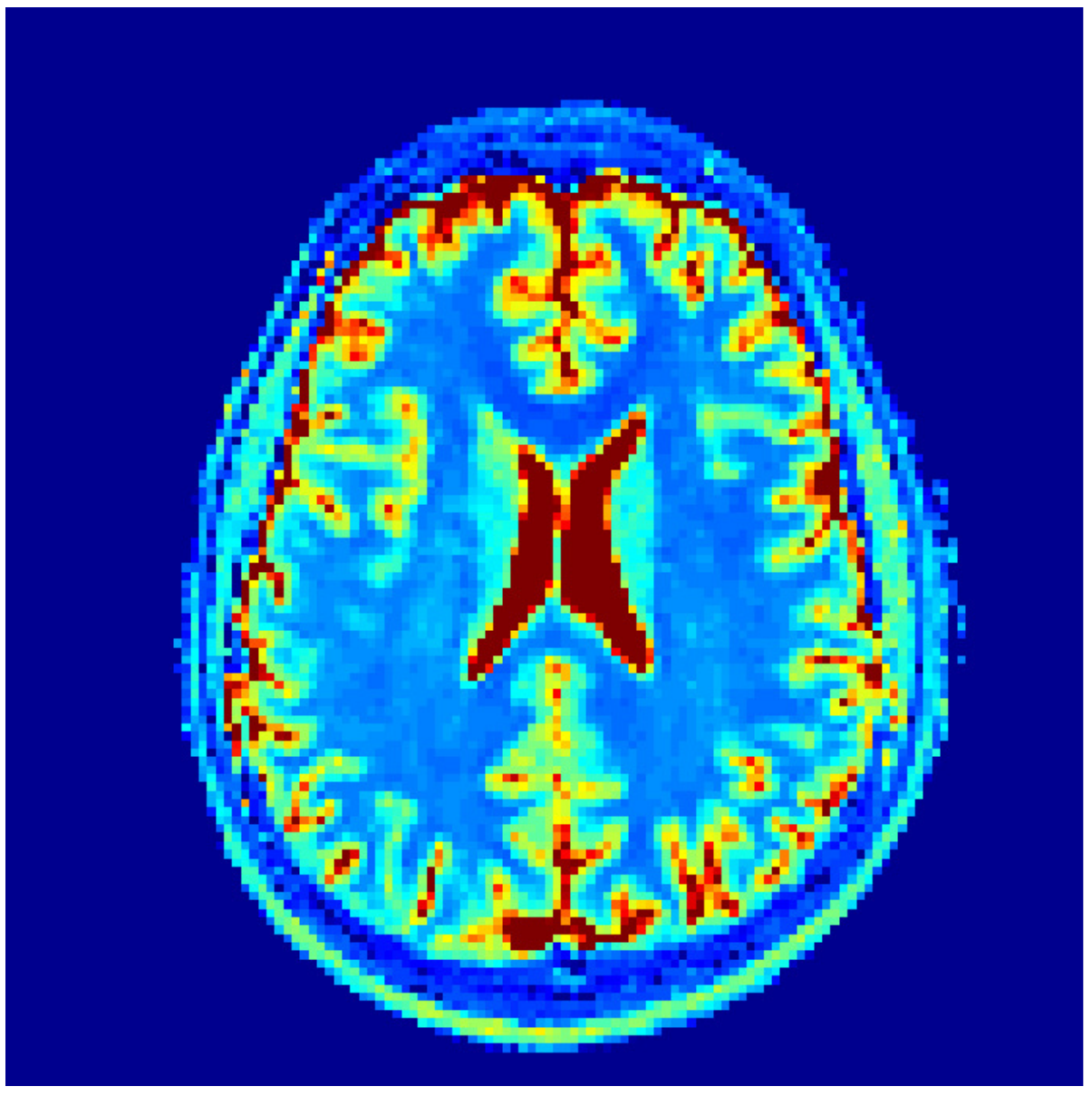}\hspace{-.1cm}
		\includegraphics[width=.162\linewidth]{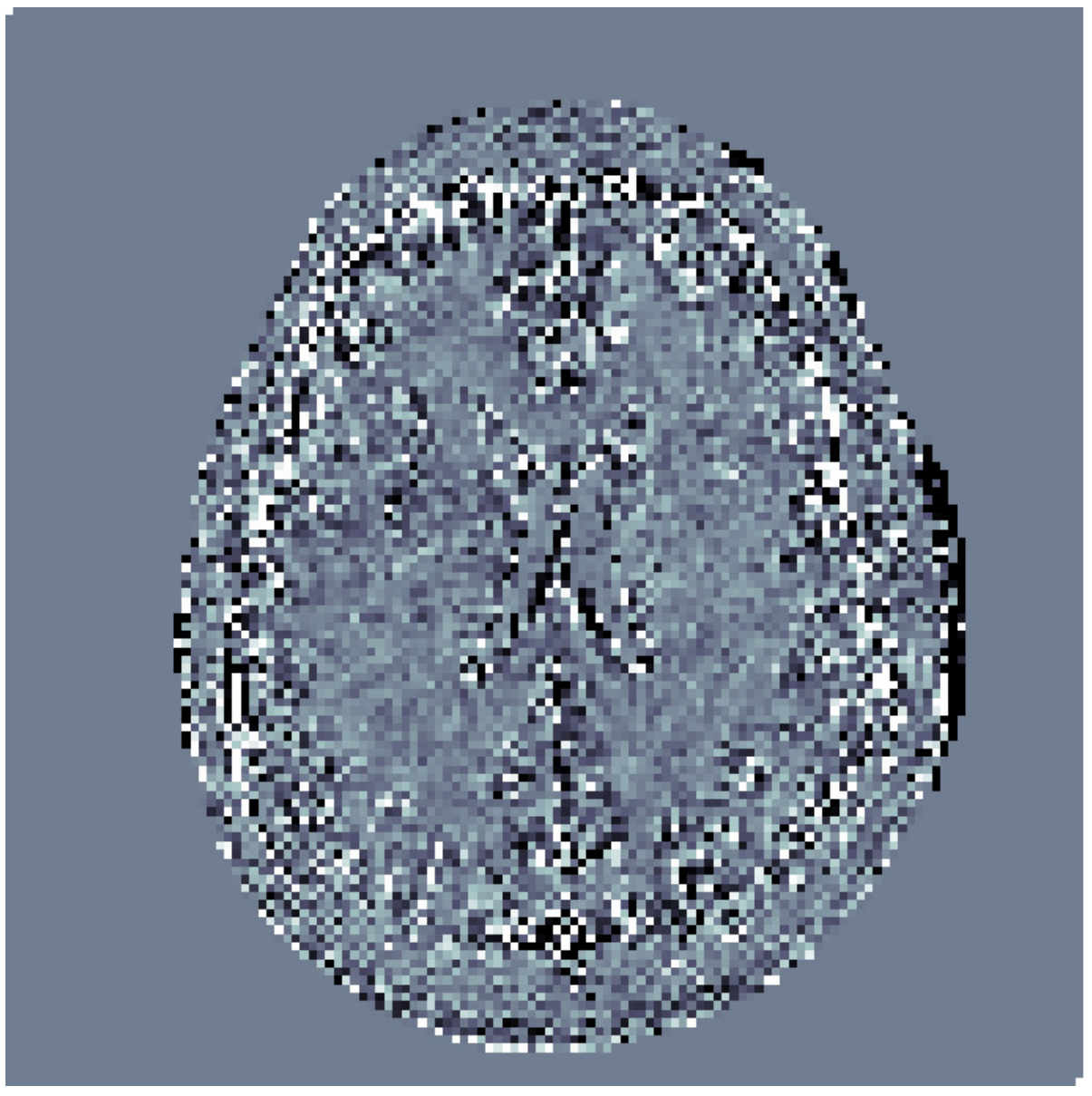}\hspace{.0cm}
		\includegraphics[width=.162\linewidth]{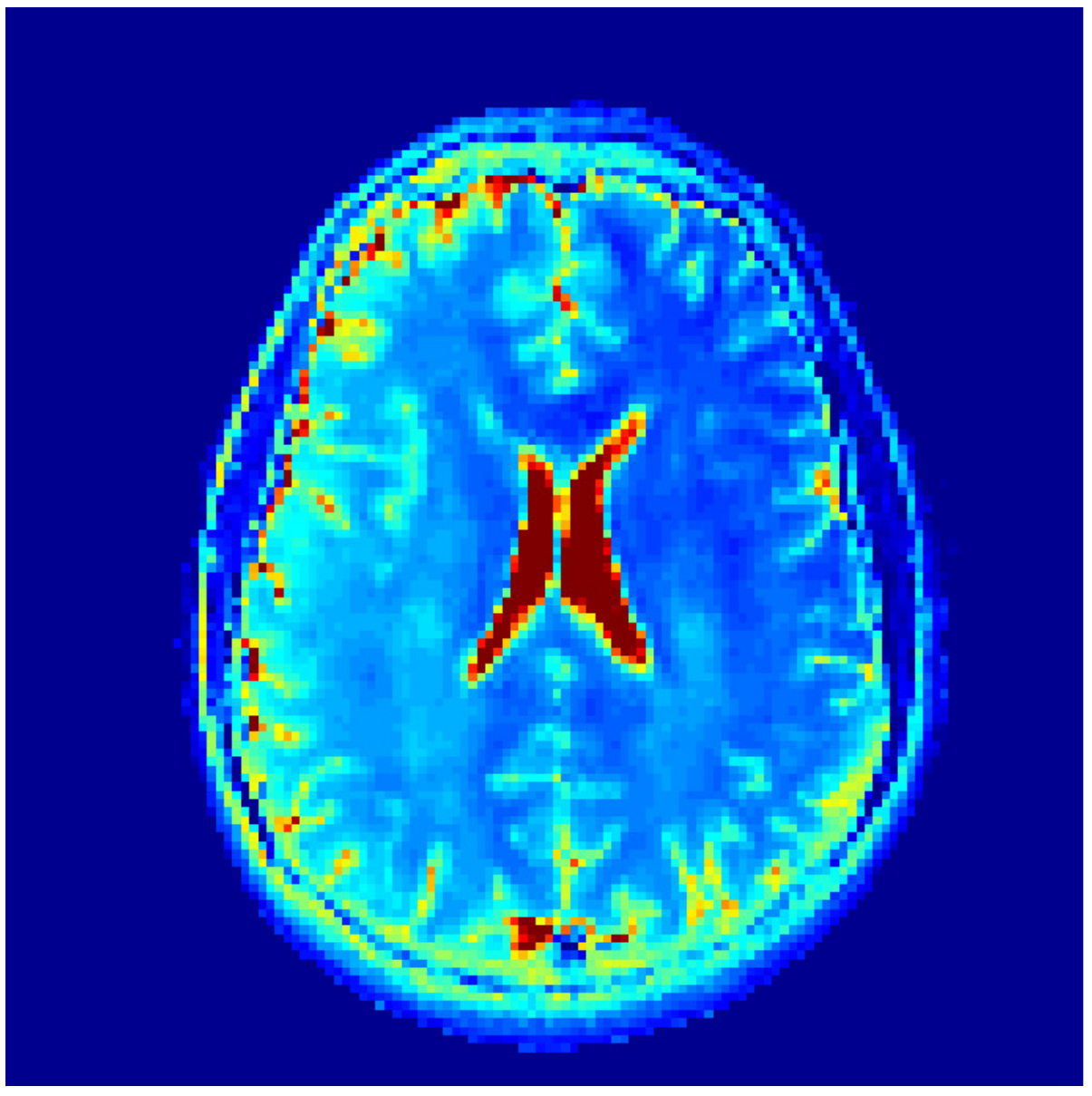}\hspace{-.1cm}
		\includegraphics[width=.162\linewidth]{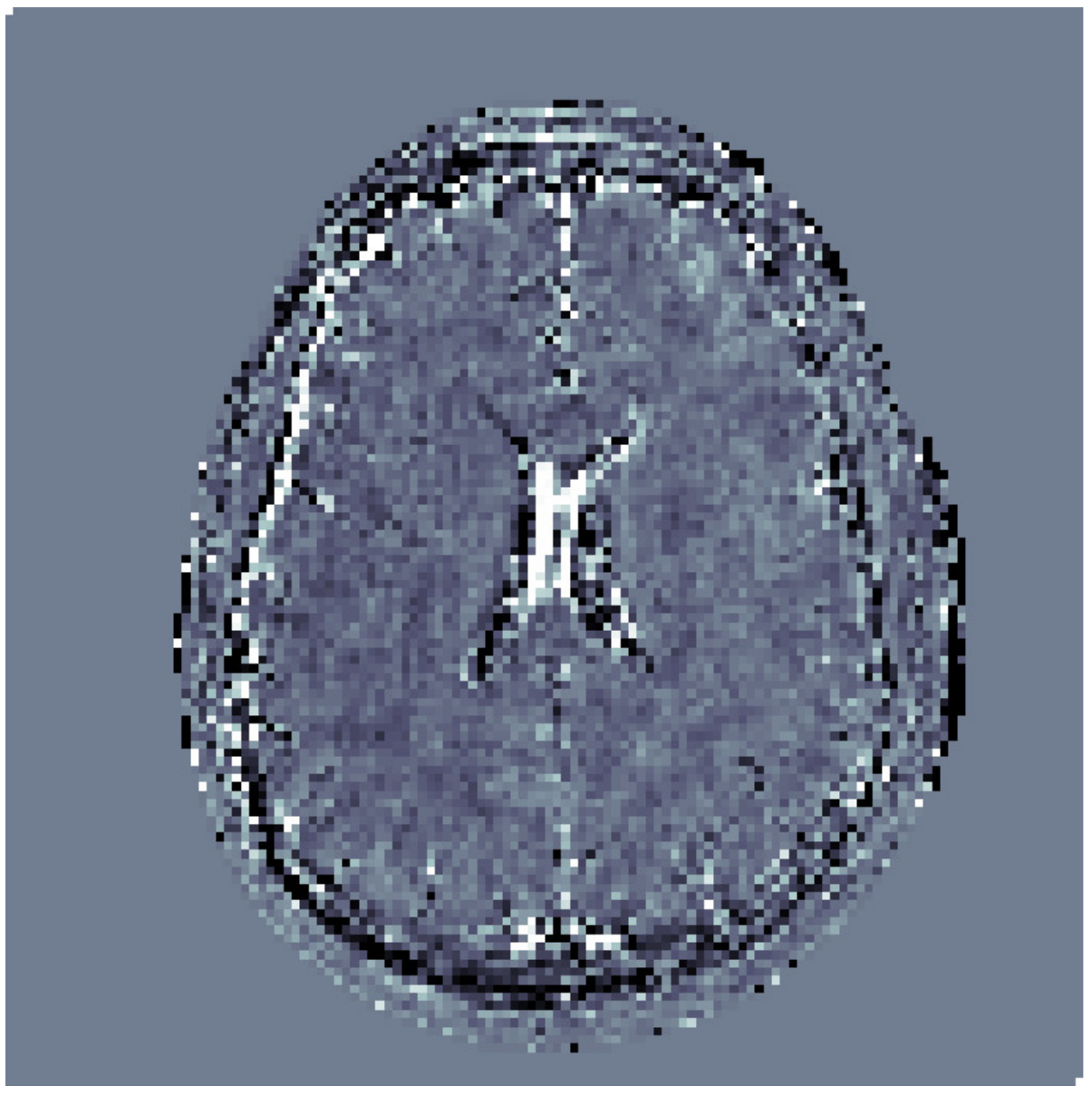}\hspace{.0cm}
		\includegraphics[width=.162\linewidth]{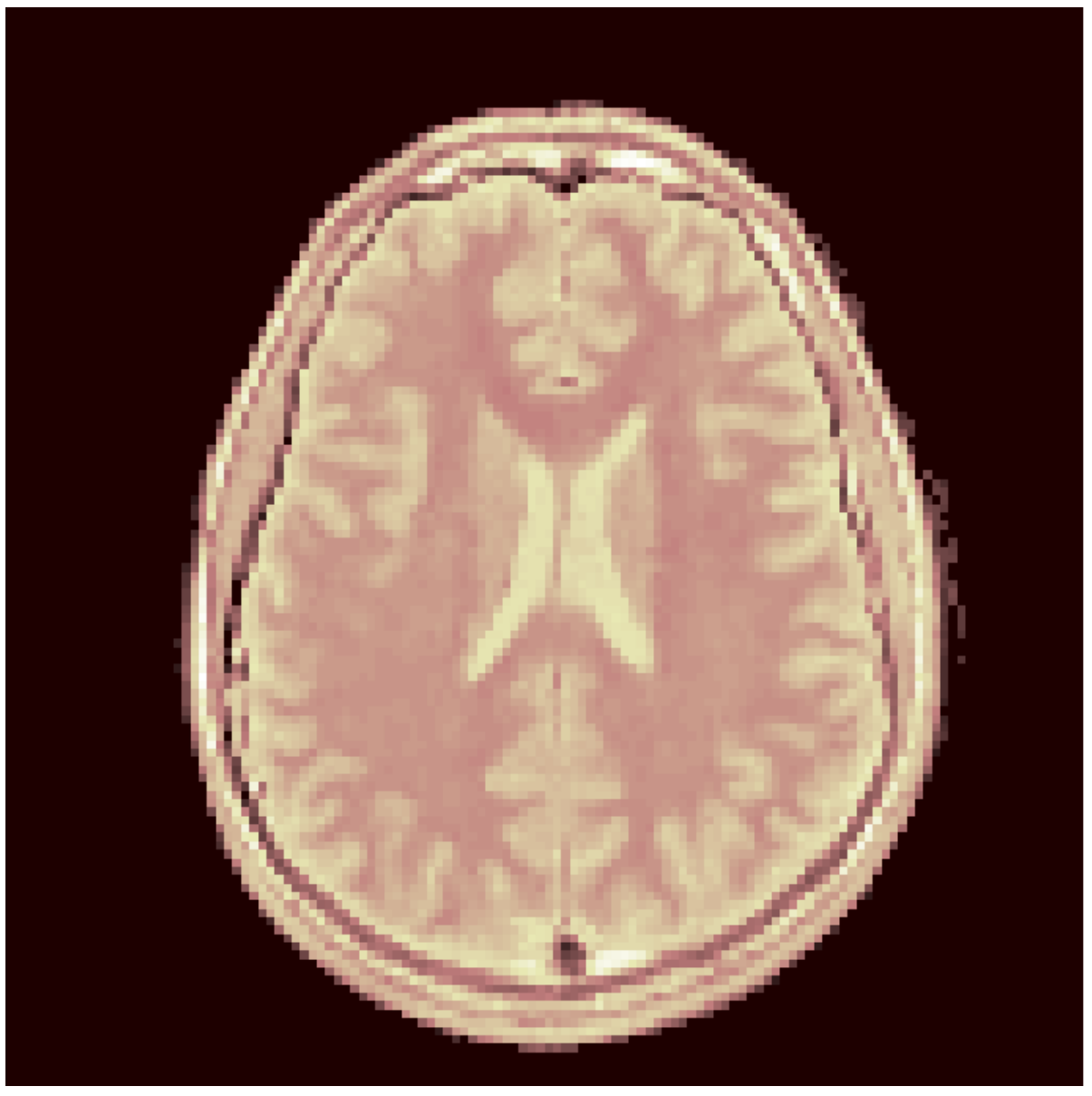}\hspace{-.1cm}
		\includegraphics[width=.162\linewidth]{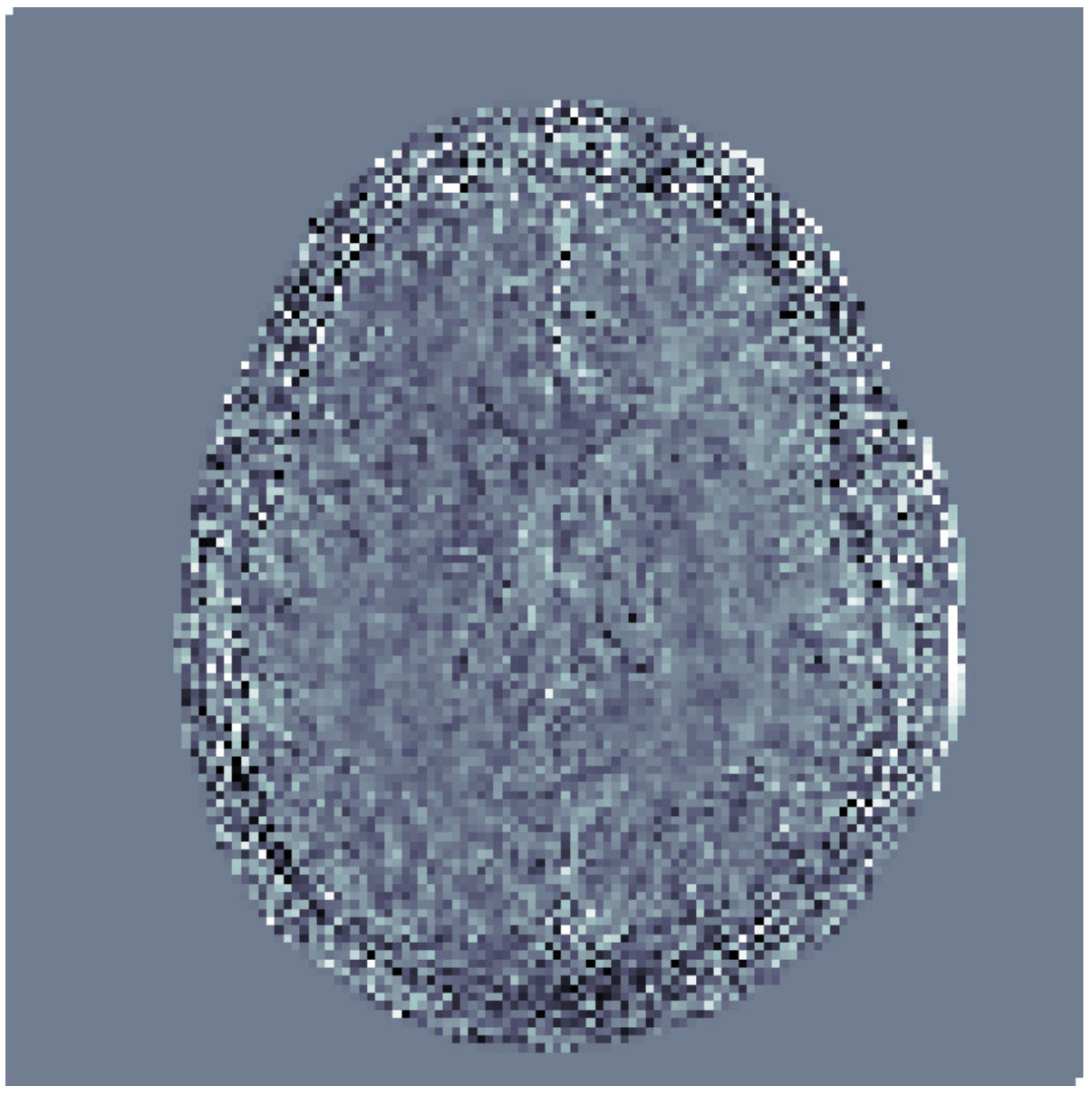}			
				\\
				\includegraphics[width=.162\linewidth]{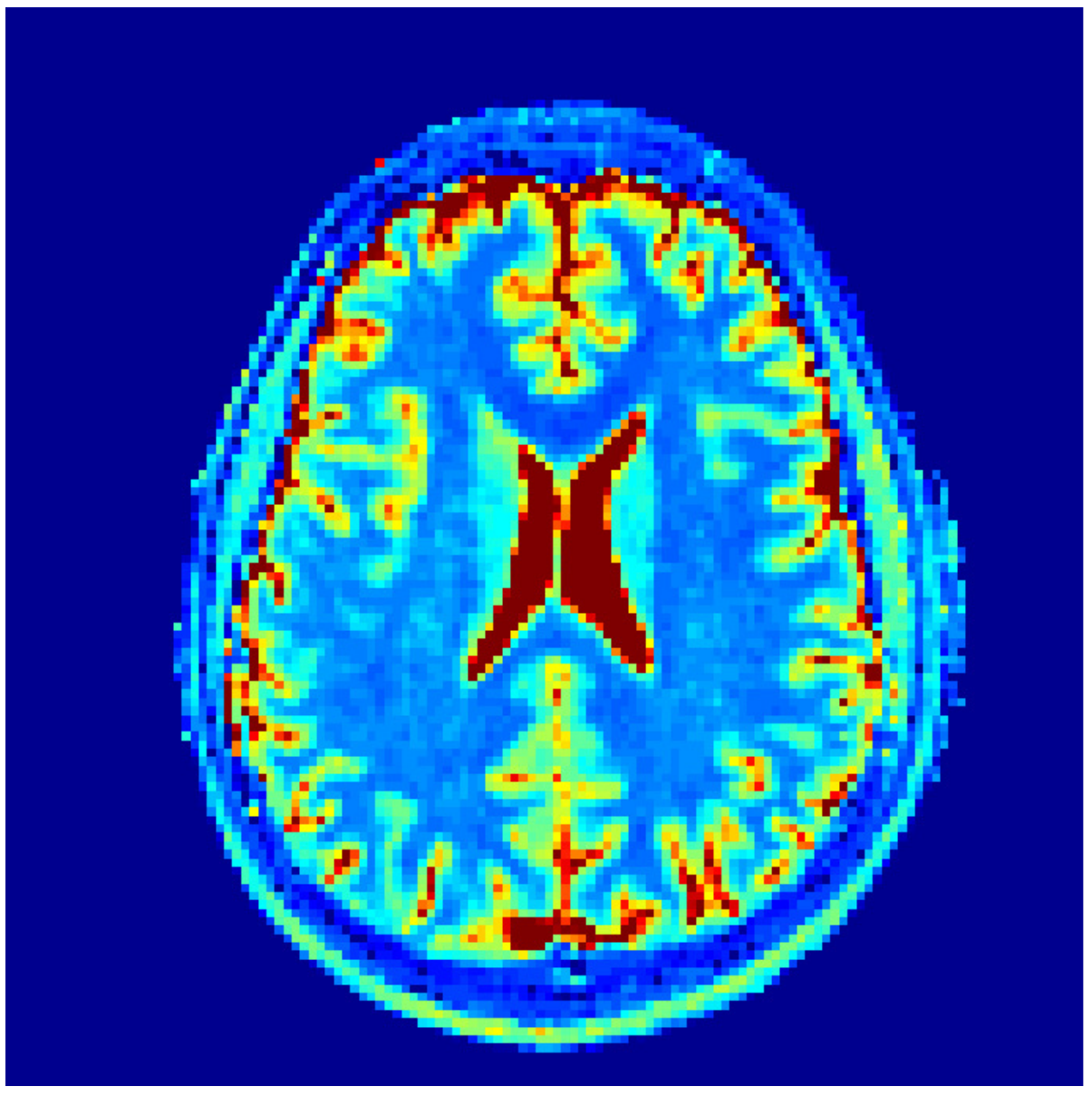}\hspace{-.1cm}
		\includegraphics[width=.162\linewidth]{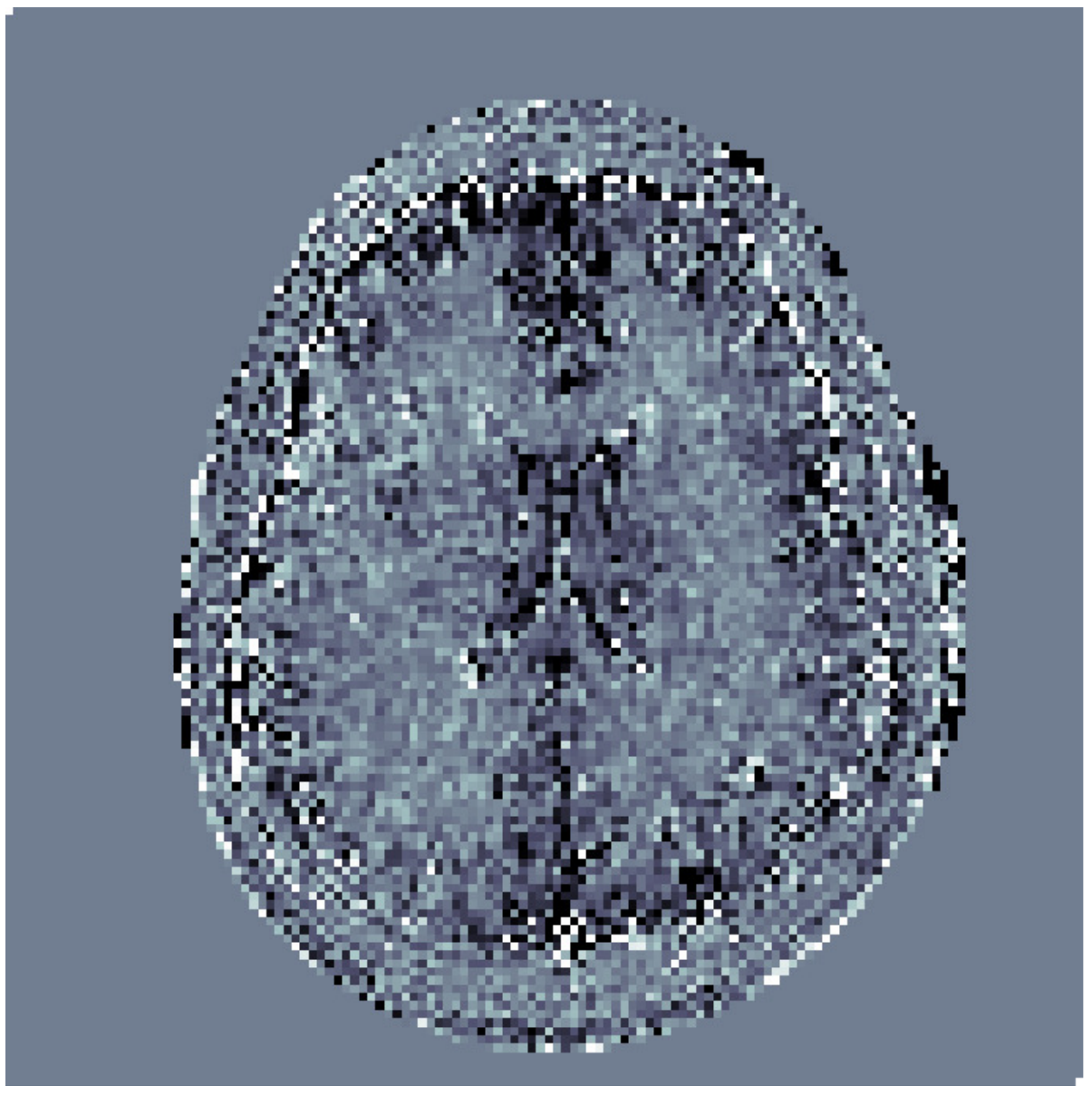}\hspace{.0cm}
		\includegraphics[width=.162\linewidth]{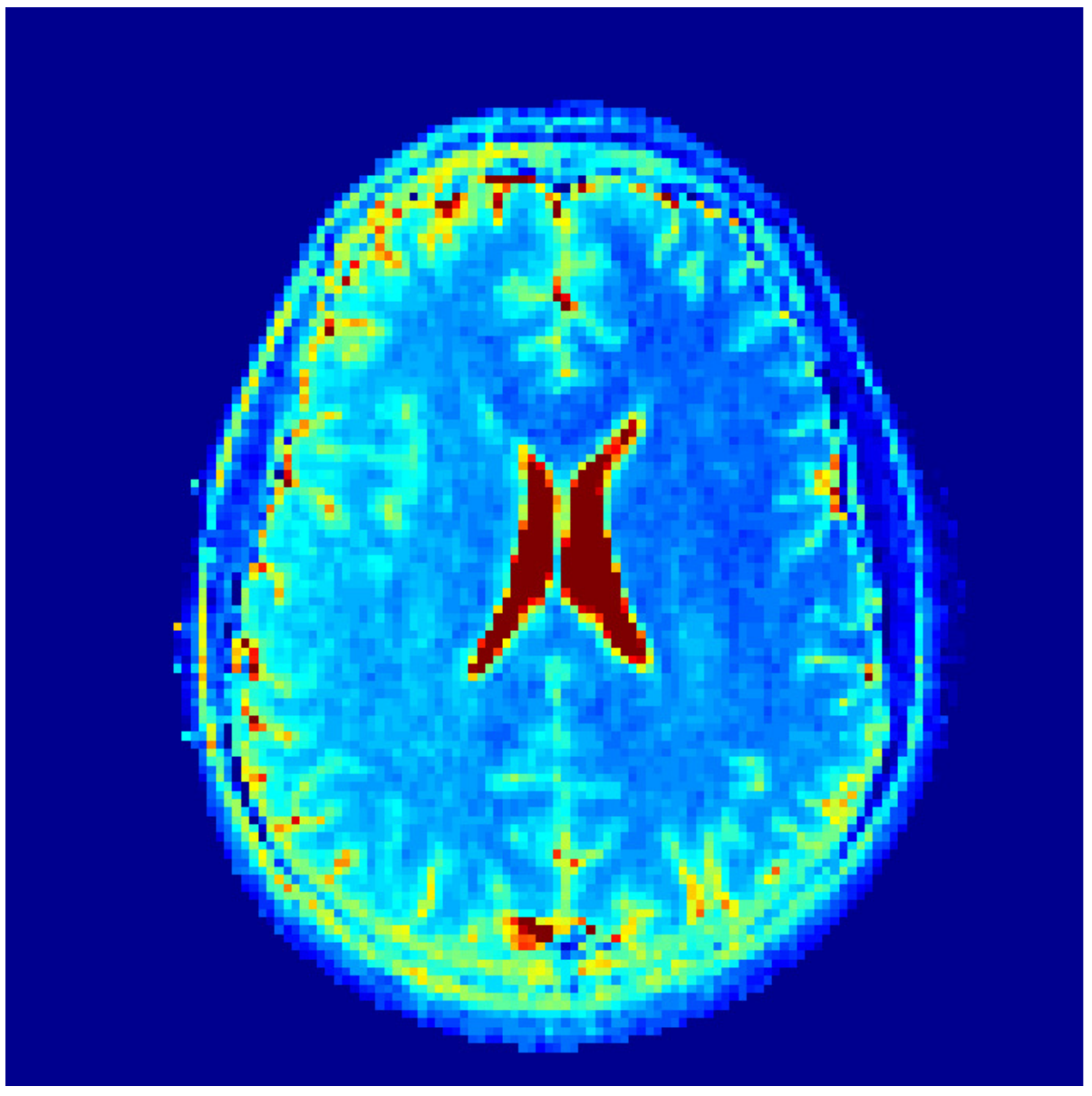}\hspace{-.1cm}
		\includegraphics[width=.162\linewidth]{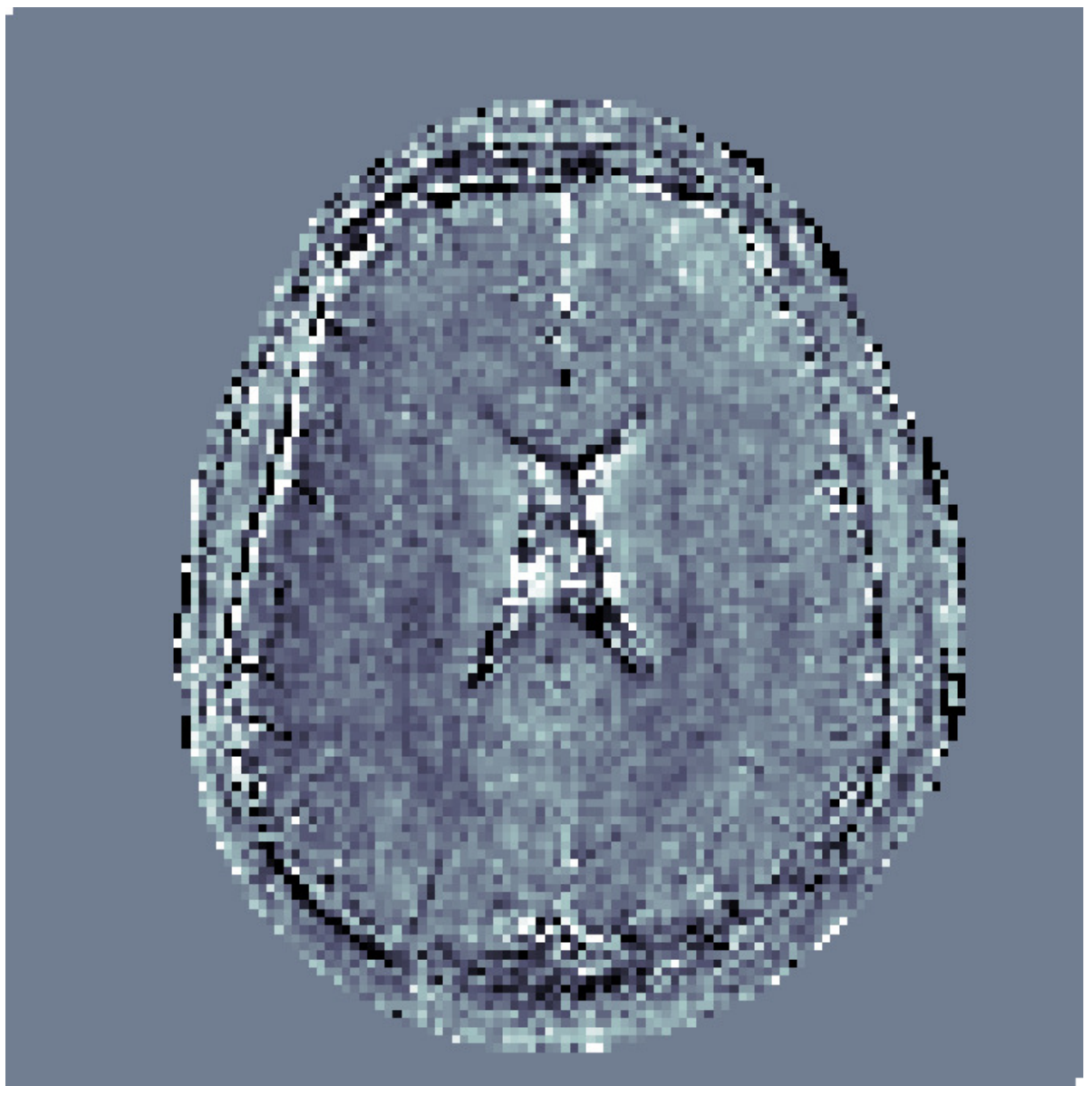}\hspace{.0cm}
		\includegraphics[width=.162\linewidth]{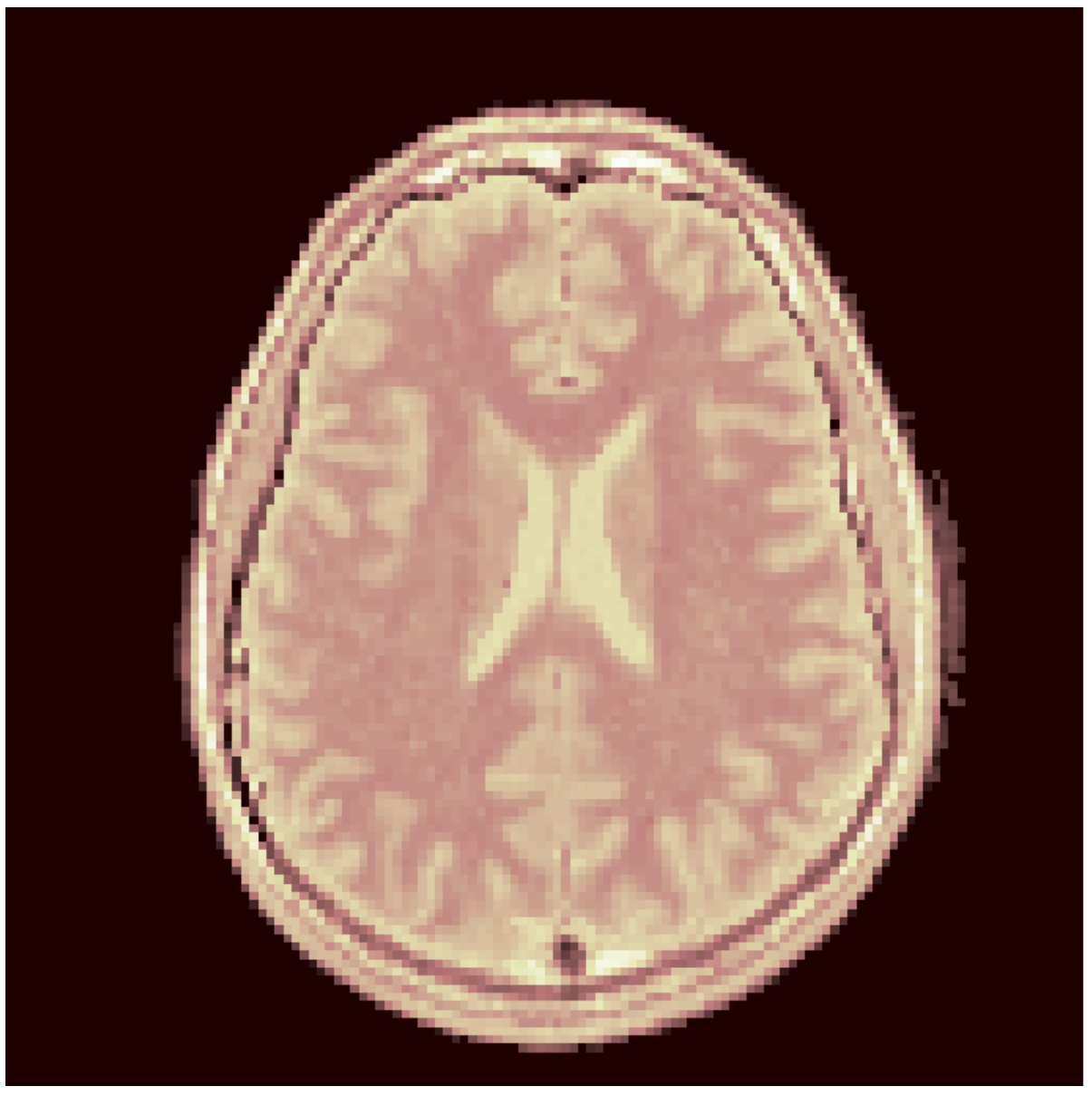}\hspace{-.1cm}
		\includegraphics[width=.162\linewidth]{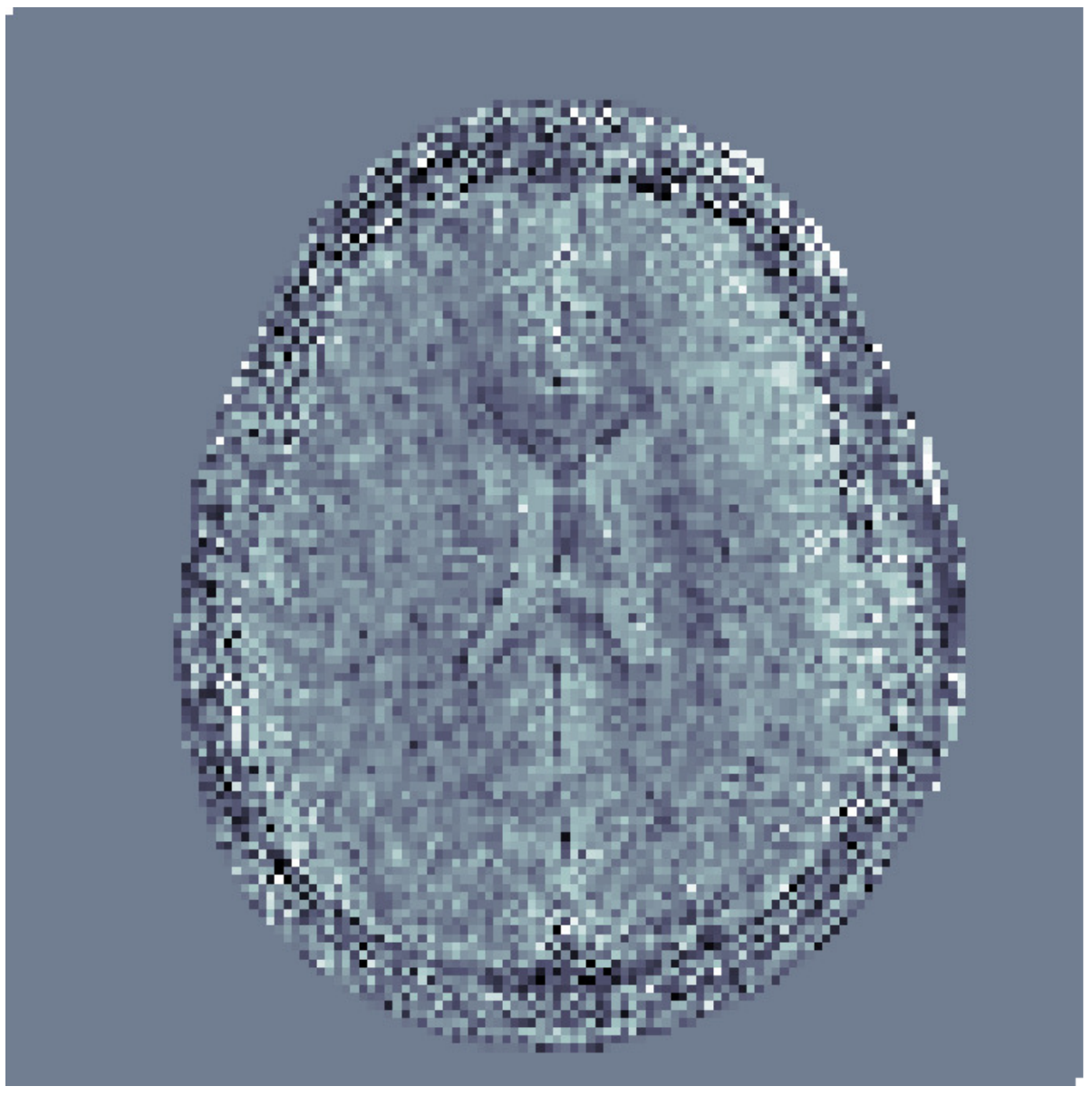}					
				\\
				\includegraphics[width=.162\linewidth]{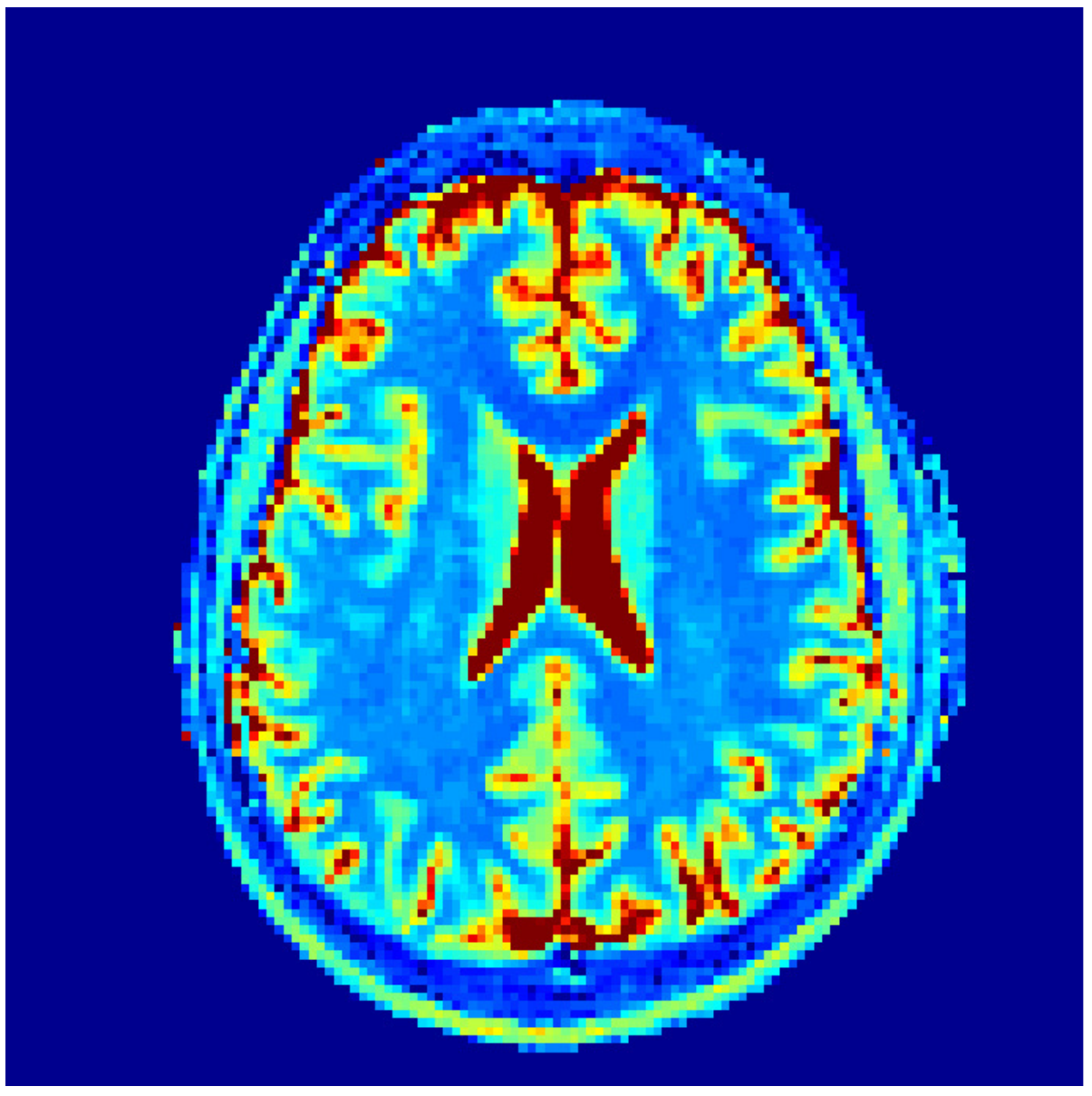}\hspace{-.1cm}
		\includegraphics[width=.162\linewidth]{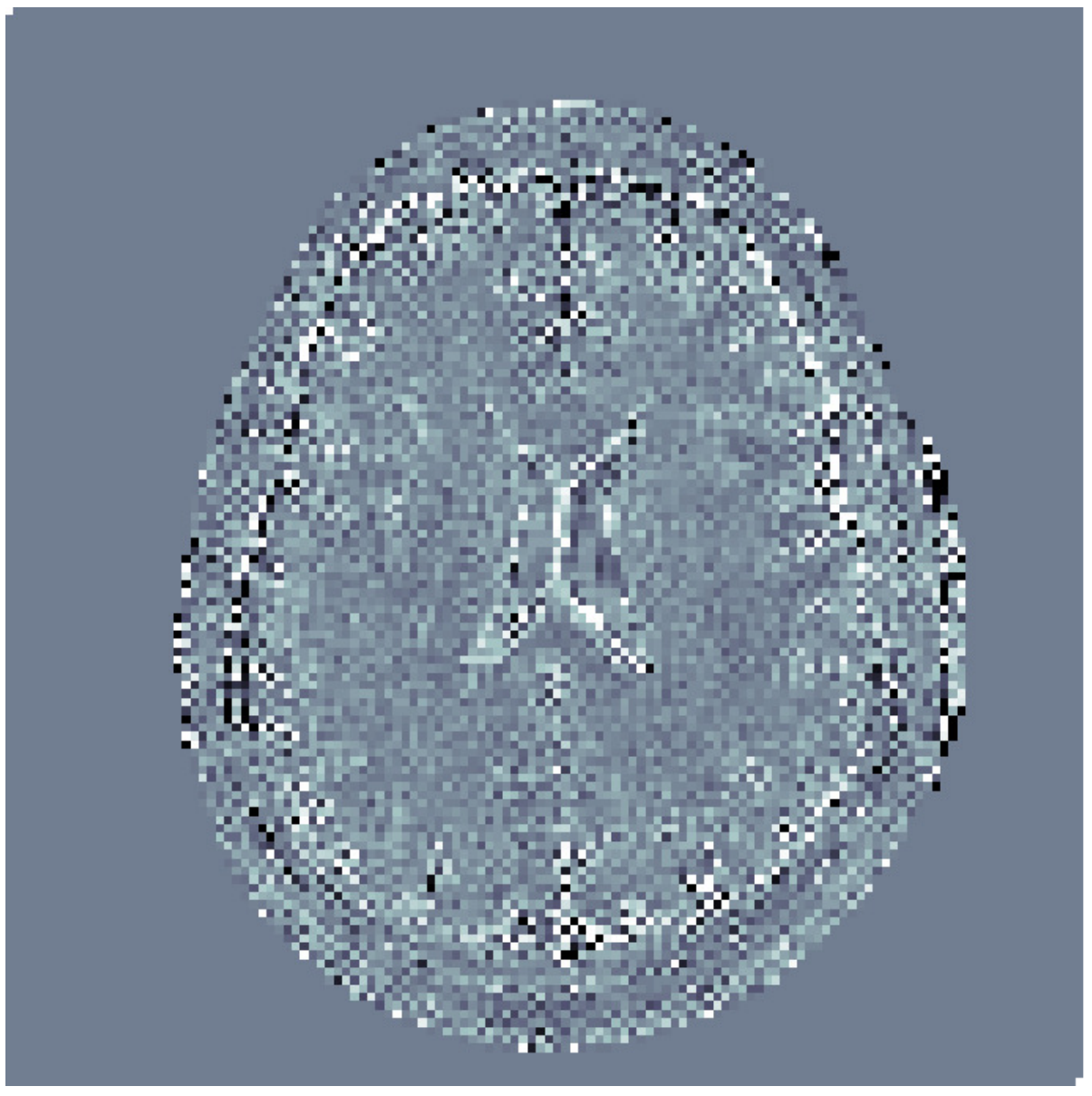}\hspace{.0cm}
		\includegraphics[width=.162\linewidth]{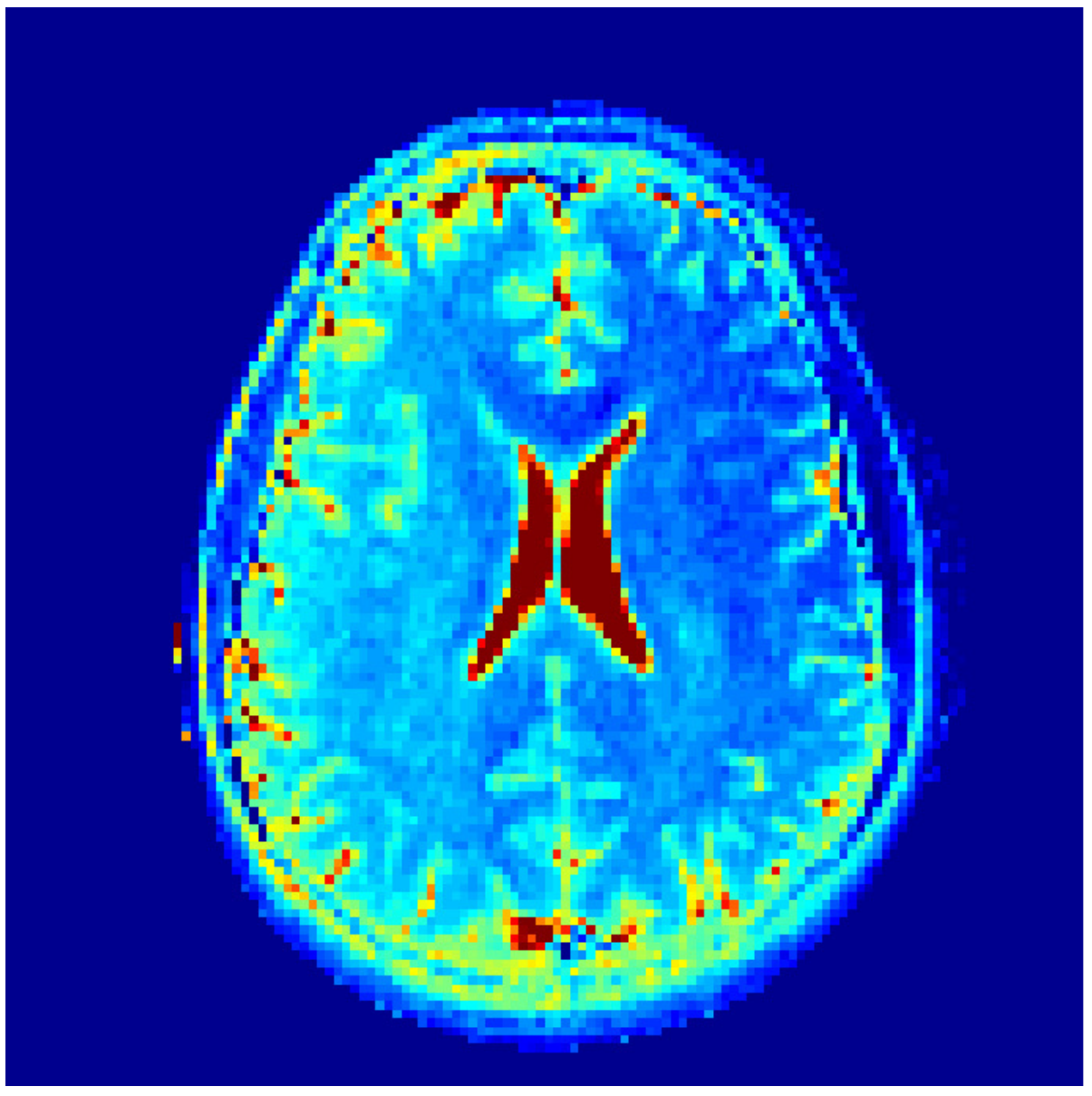}\hspace{-.1cm}
		\includegraphics[width=.162\linewidth]{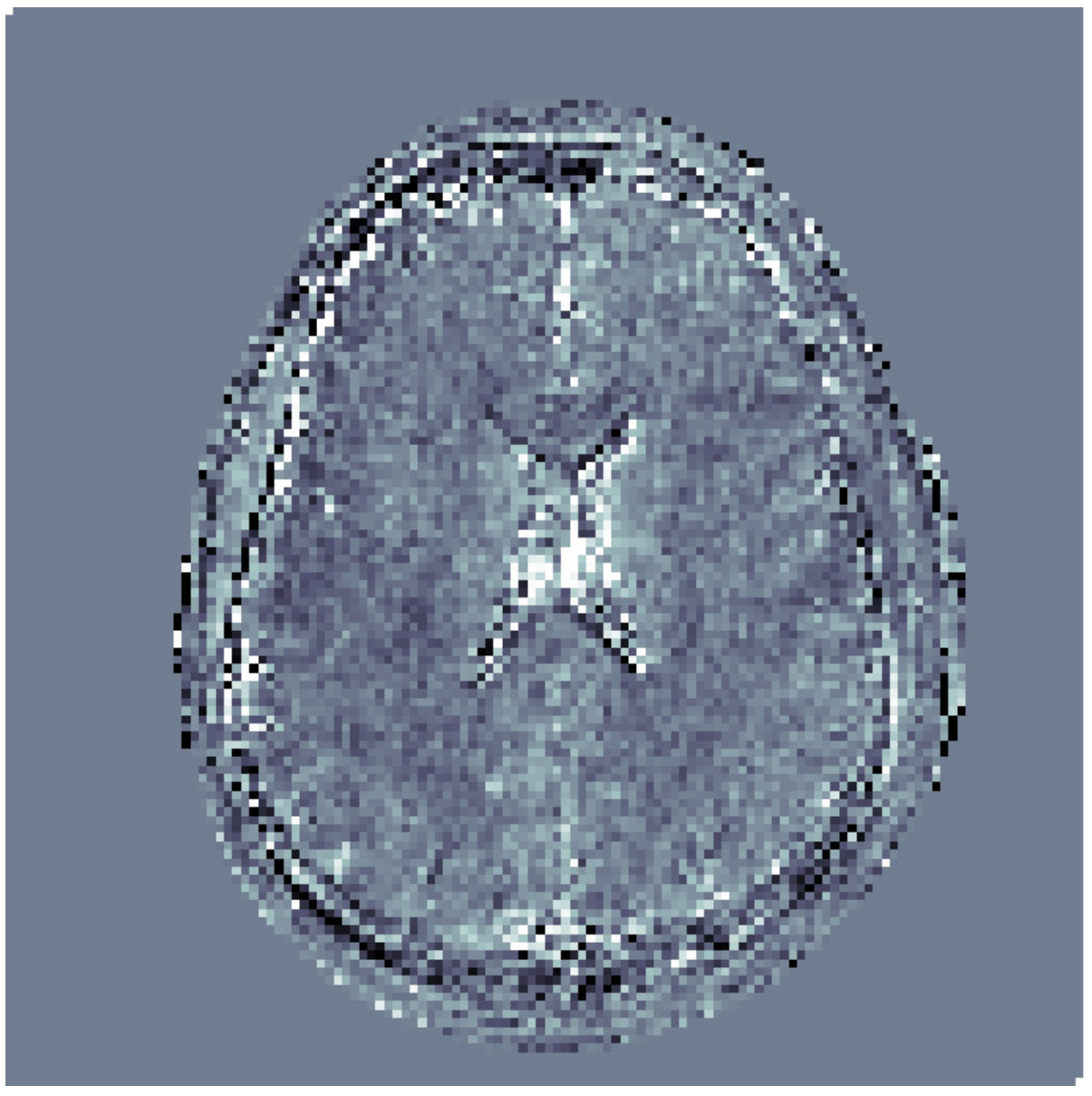}\hspace{.0cm}
		\includegraphics[width=.162\linewidth]{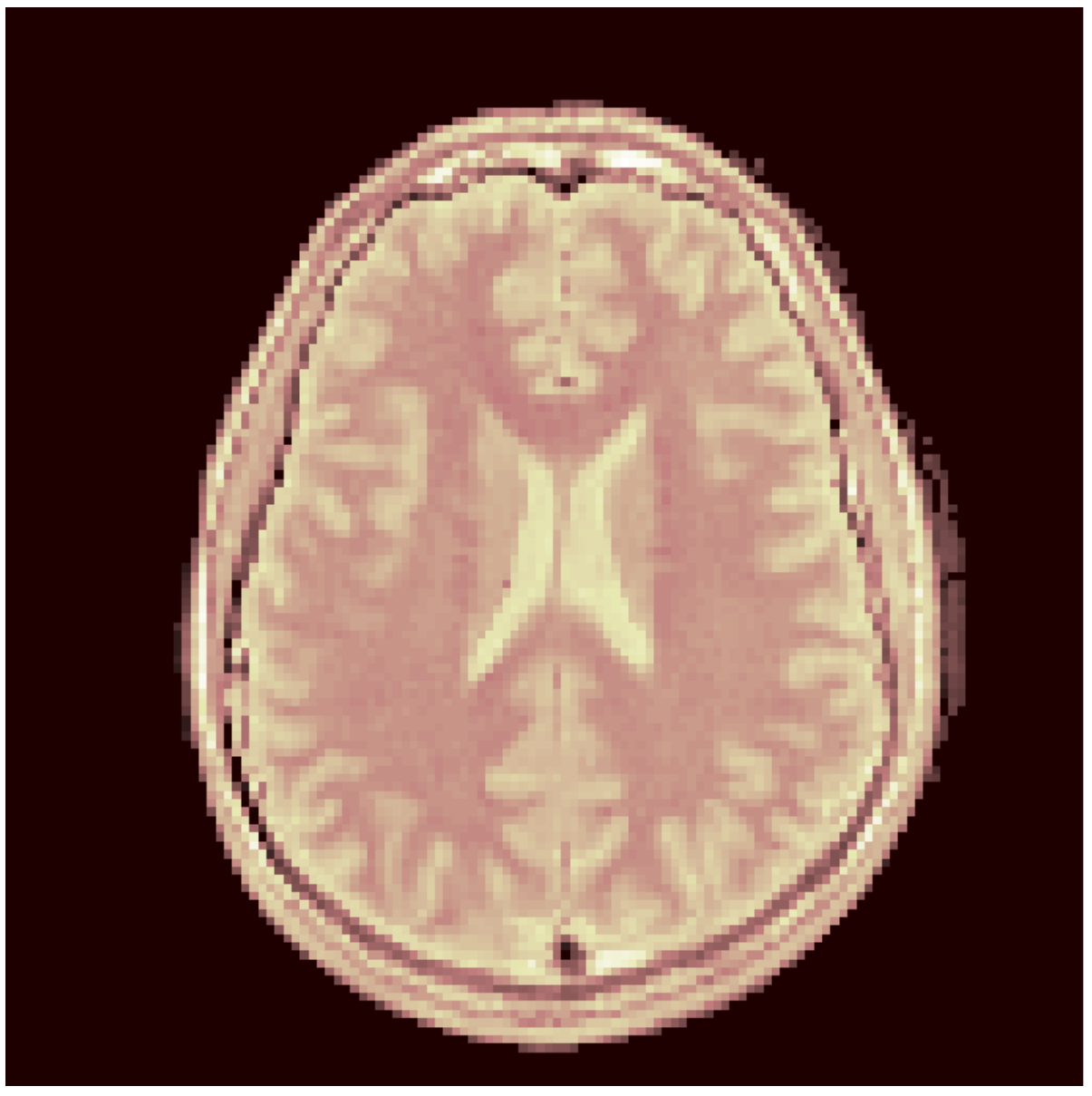}\hspace{-.1cm}
		\includegraphics[width=.162\linewidth]{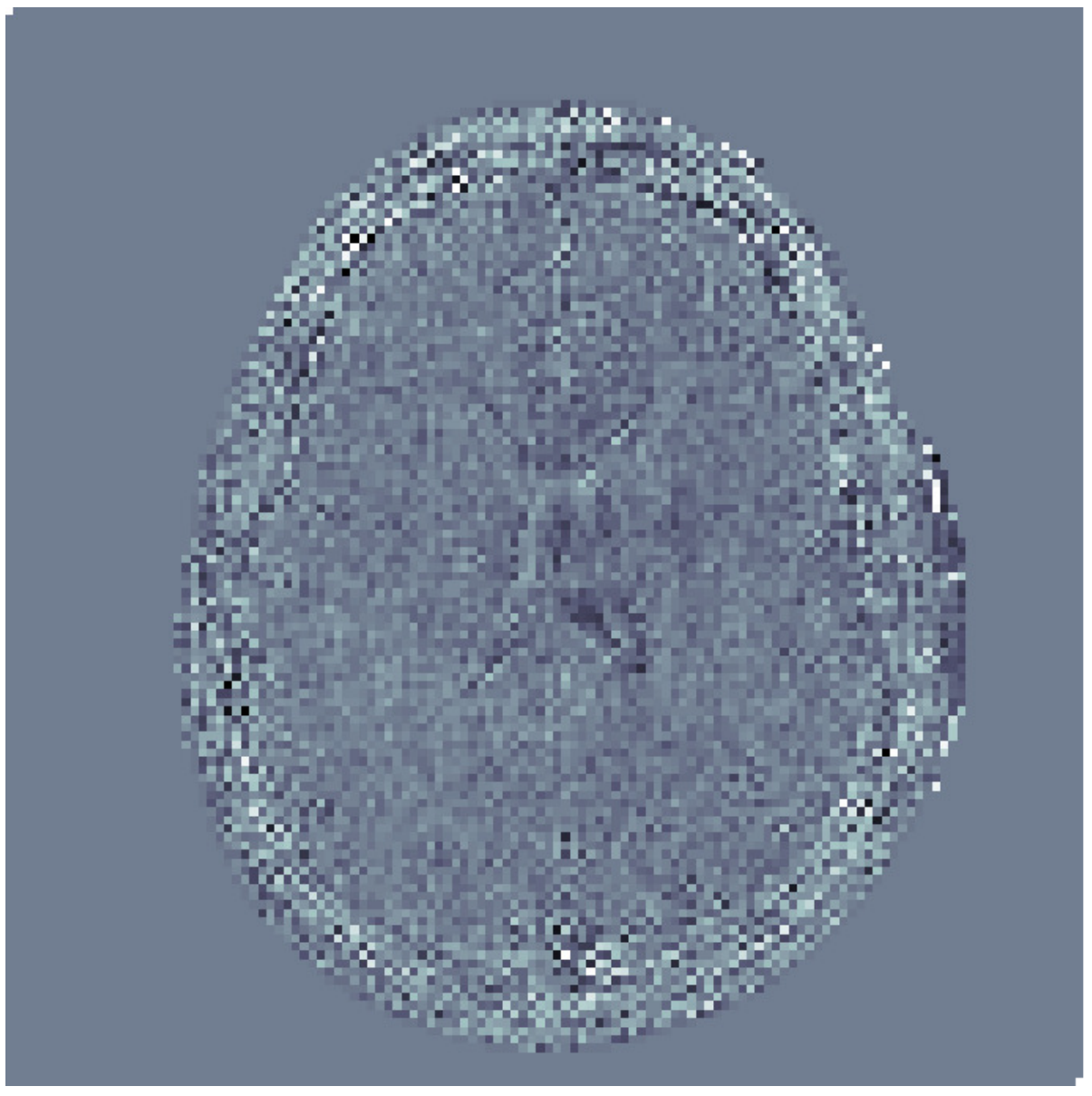}			
				\\
		\includegraphics[width=.162\linewidth]{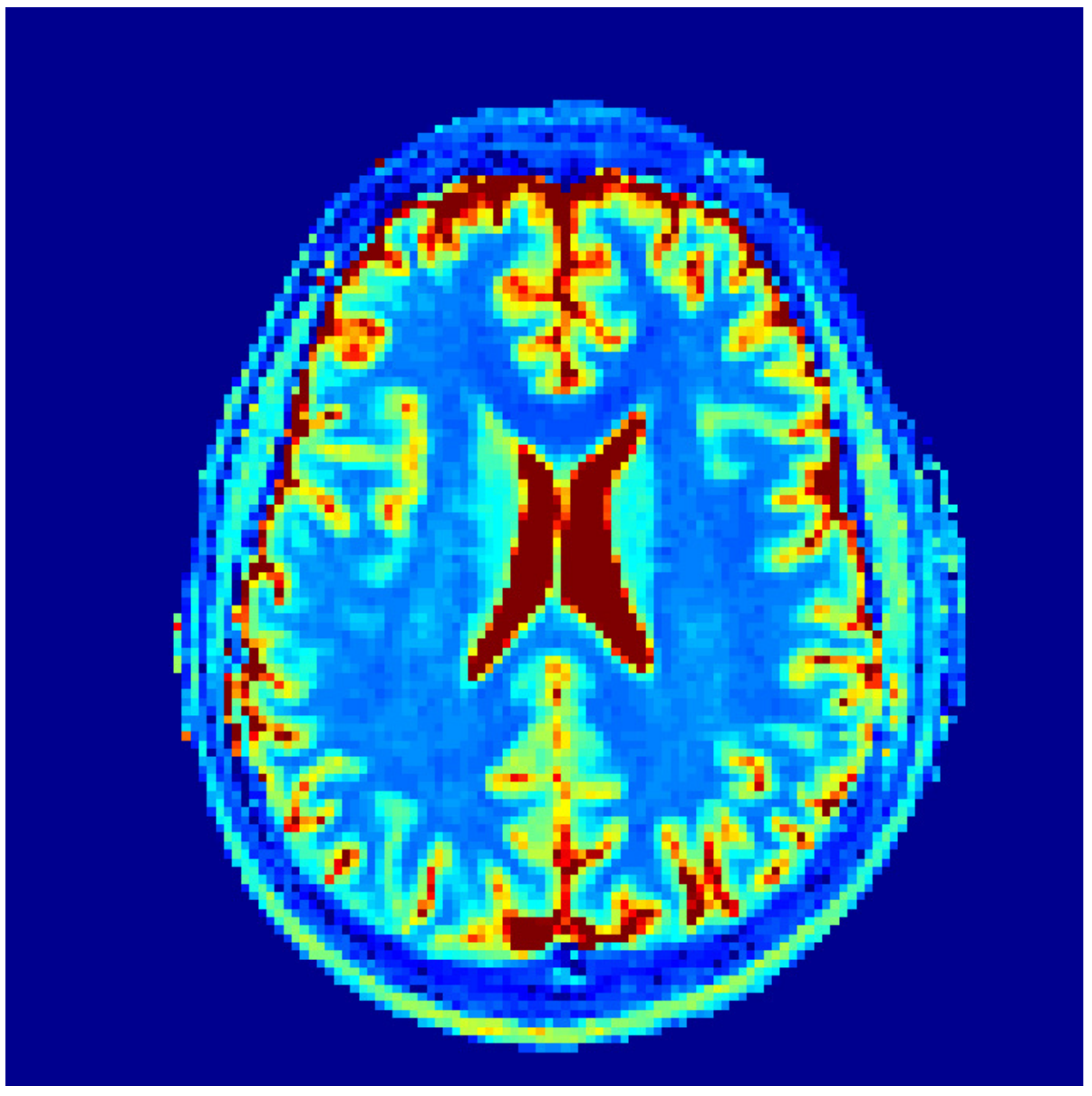}\hspace{-.1cm}
		\includegraphics[width=.162\linewidth]{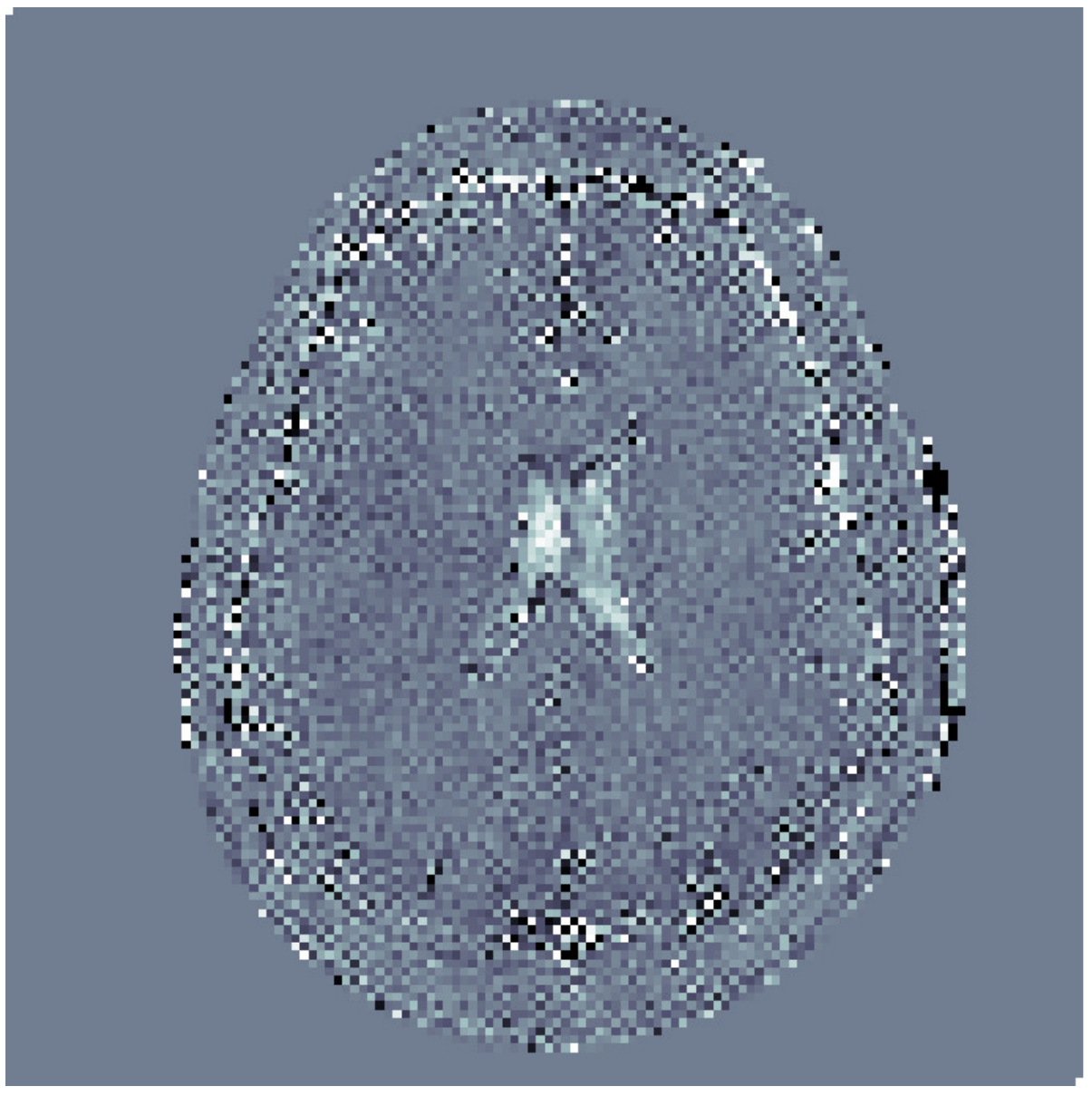}\hspace{.0cm}
		\includegraphics[width=.162\linewidth]{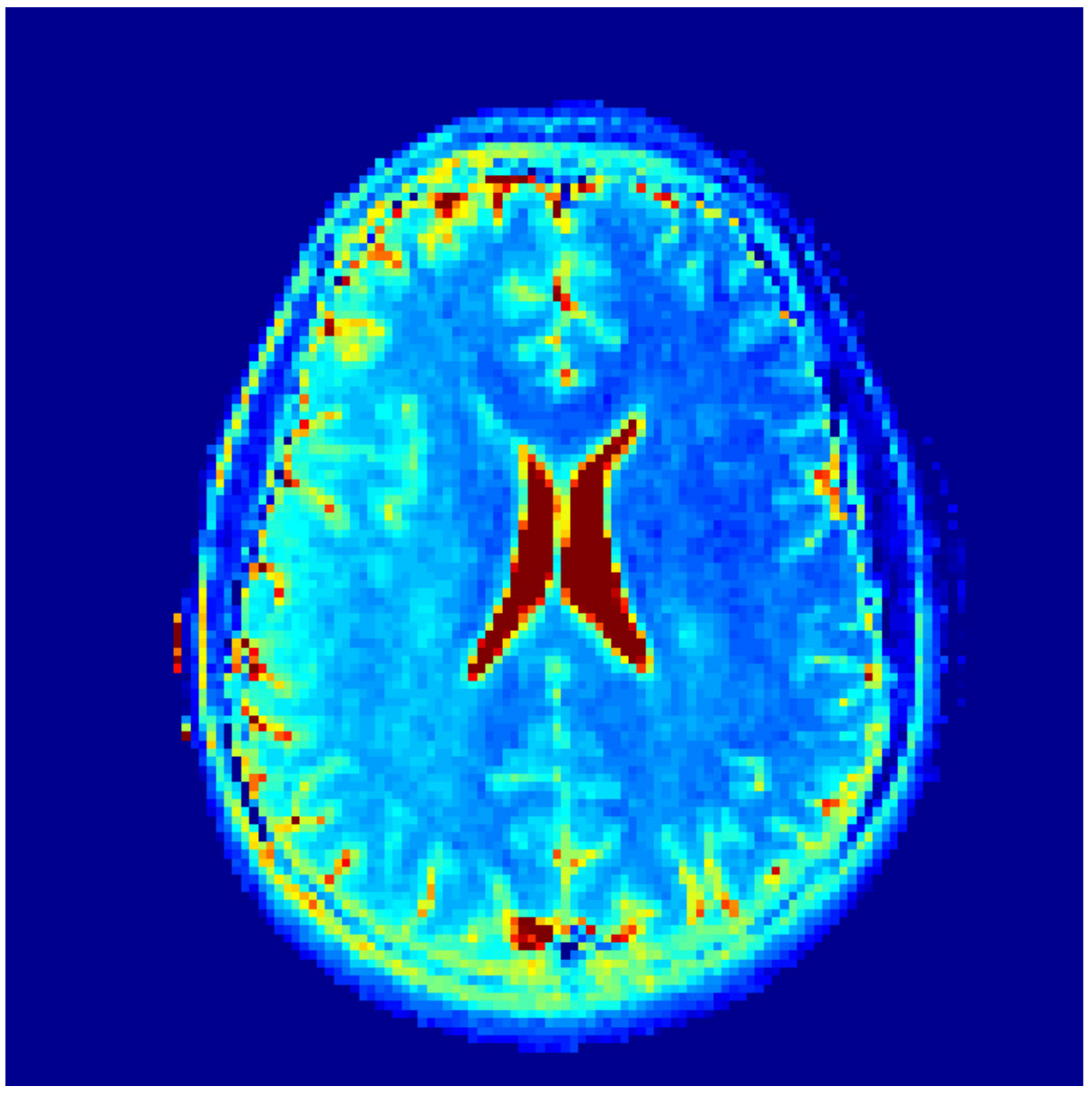}\hspace{-.1cm}
		\includegraphics[width=.162\linewidth]{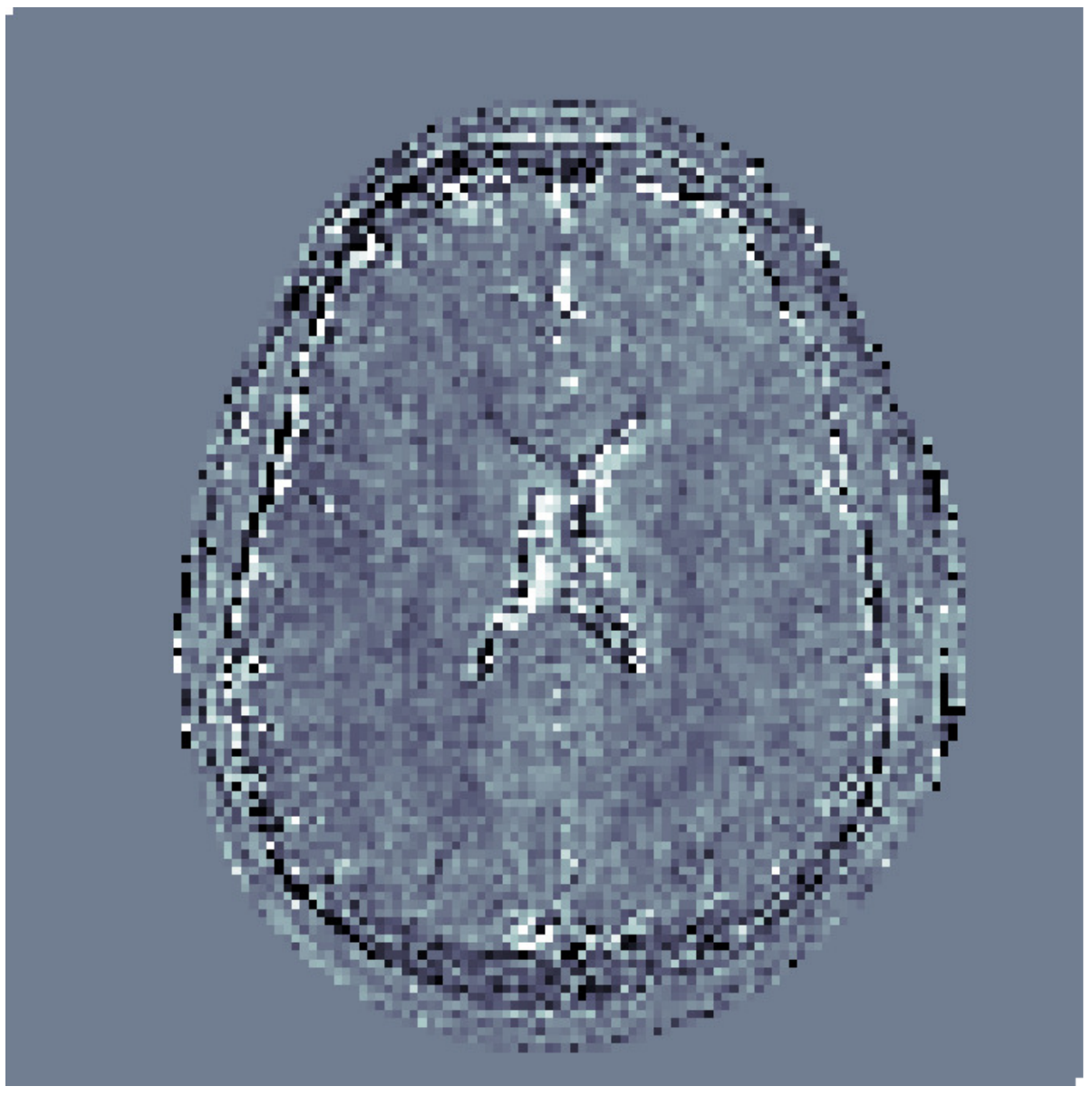}\hspace{.0cm}
		\includegraphics[width=.162\linewidth]{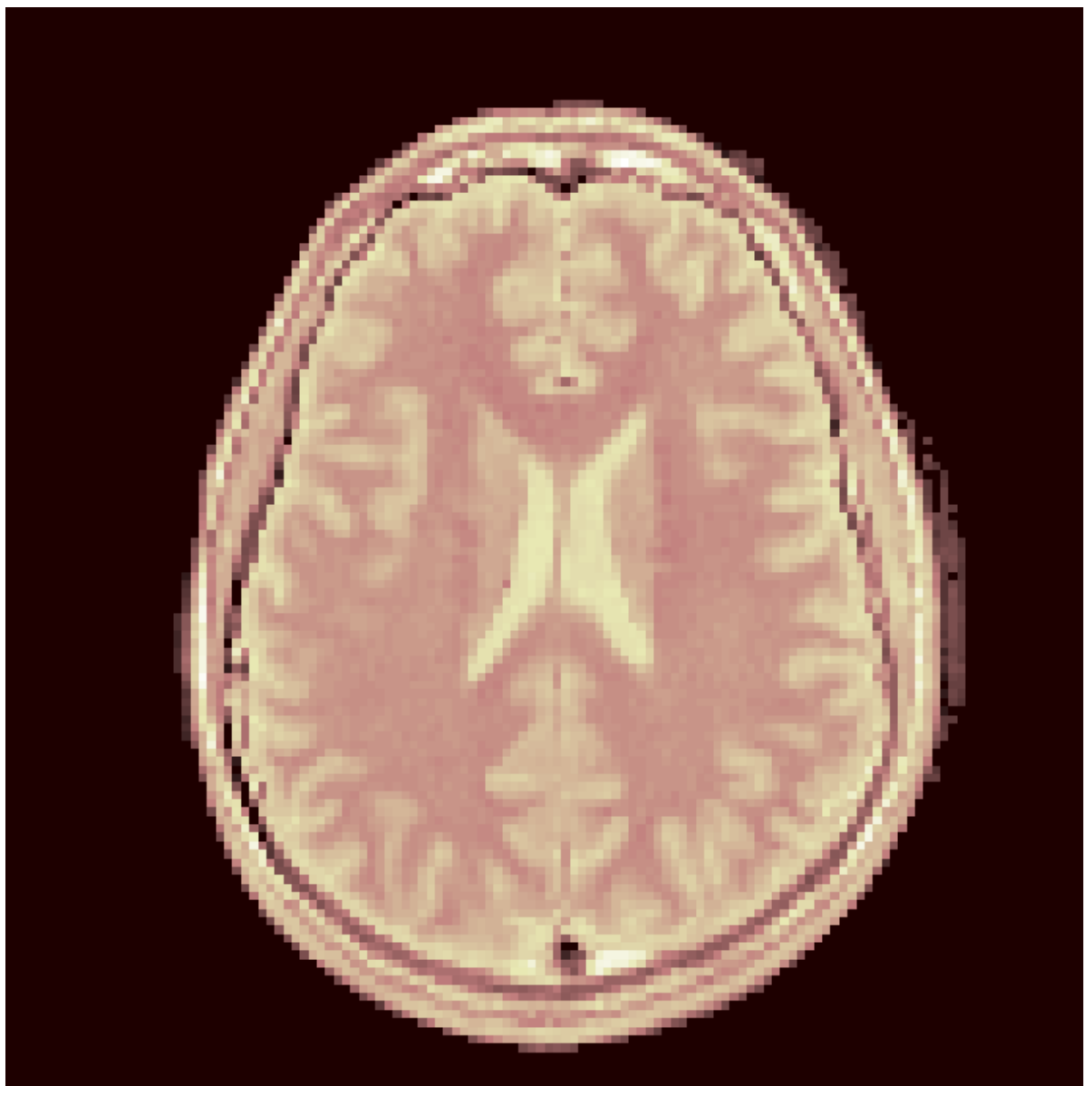}\hspace{-.1cm}
		\includegraphics[width=.162\linewidth]{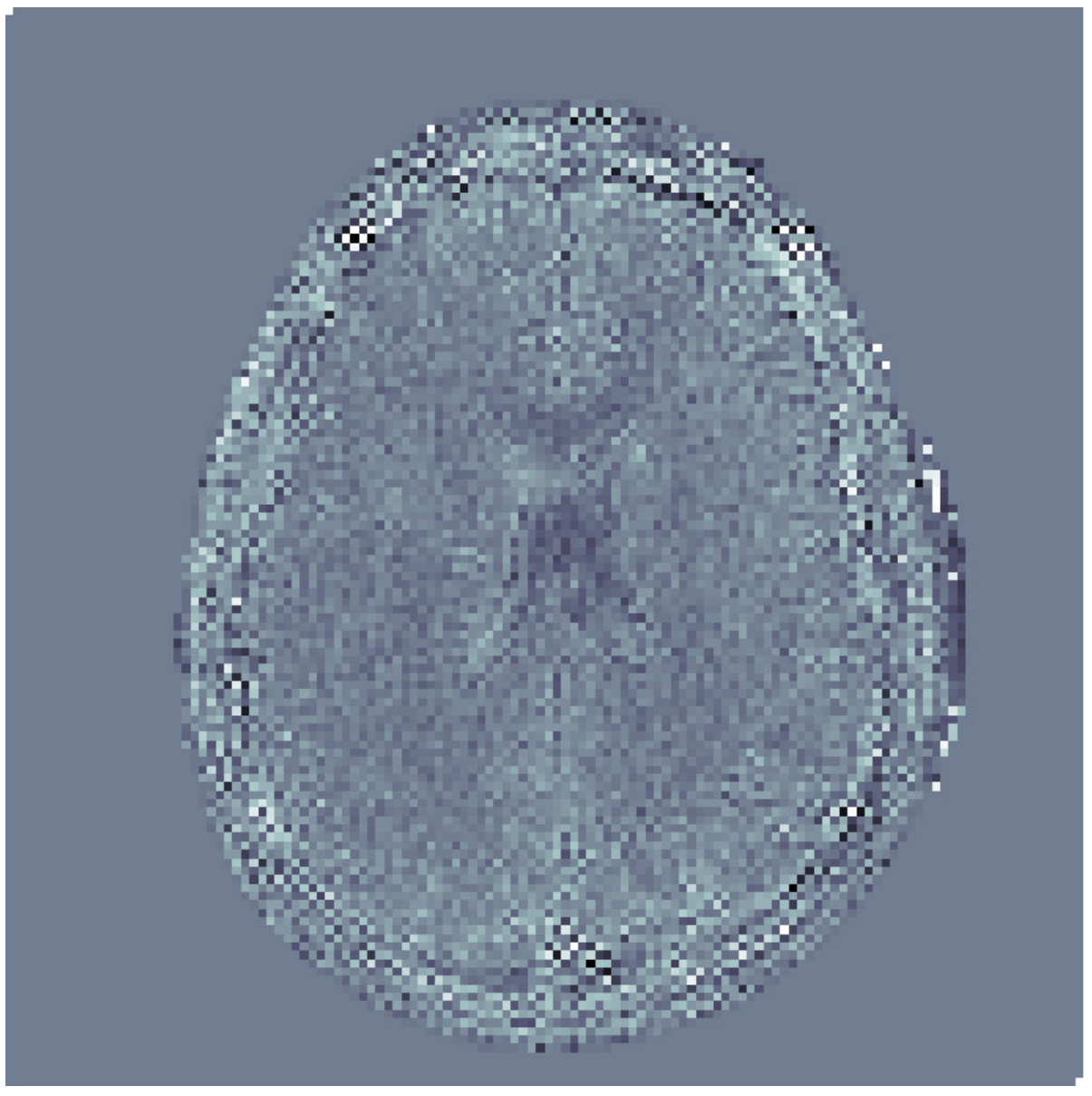}\vspace{-.3cm}			
		\\
		\includegraphics[trim= -10 50 -20 720, clip,width=.17\linewidth]{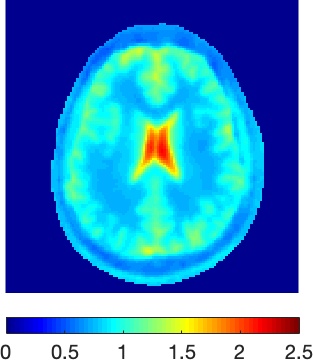}\hspace{-.2cm}
		\includegraphics[trim= -10 50 -20 700, clip,width=.17\linewidth]{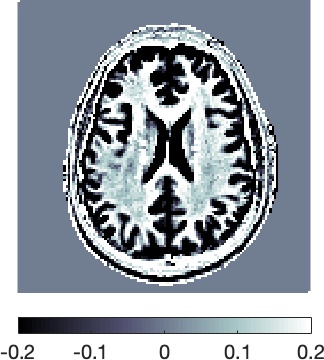}\hspace{-.2cm}
		\includegraphics[trim= -10 50 -20 700, clip,width=.17\linewidth]{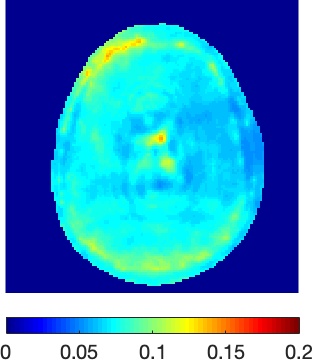}\hspace{-.2cm}
		\includegraphics[trim= -10 50 -20 700, clip,width=.17\linewidth]{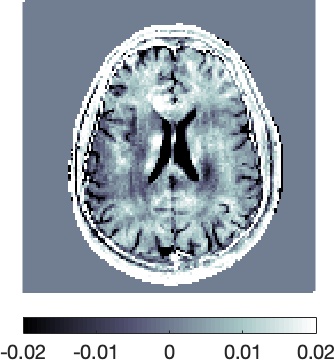}\hspace{-.2cm}
		\includegraphics[trim= -10 50 -20 700, clip,width=.17\linewidth]{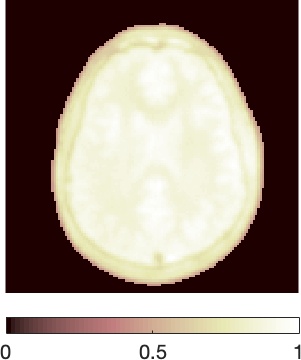}\hspace{-.2cm}
		\includegraphics[trim= -10 50 -20 700, clip,width=.17\linewidth]{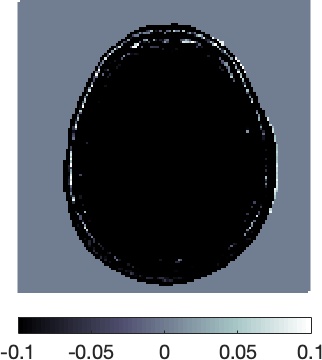}\vspace{.3cm}
		\\
		\footnotesize{T1 (sec) \hspace{0.7cm} T1 error  \hspace{.7cm} T2 (sec) \hspace{.7cm} T2 error \hspace{.7cm} PD (a.u.) \hspace{.7cm} PD error}
		\\
		\caption{{A slice of the true T1, T2 and PD maps acquired by the gold standard MAGIC  (top), and the corresponding MRF reconstructions using (from top to bottom) FGM, BLIP+FGM, MRFCNN, SCQ, encoder alone $\G$, and the \proxnetB with $T = 2$ and $T=5$ algorithms.
%Figure (c) displays the ground truth maps used for simulations.
}  \label{fig:spiral_recon}}
\end{minipage}}
\end{figure*}
\clearpage

\subsection{Results and discussions}
Table~\ref{tab:testdata1} and Figure~\ref{fig:spiral_recon} compare the performances of the MRF baselines against our proposed \proxnet\ using $T=2$ and 5 recurrent iterations. We also include inference results using the proposed \textit{encoder alone} $\G$, without proximal iterations.  Reconstruction performances were measured by the Normalised RMSE $=\frac{\|T1-T1^{GT}\|}{\|T1^{GT}\|}$,  MAE $=|T1-T1^{GT}|$, Structural Similarity Index Metric (SSIM) \cite{wang2004image}, the required storage for the MRF dictionary (in DM methods) or the networks, and the algorithm runtimes averaged over the test image slices.

The non-iterative FGM results in incorrect maps due to the severe under-sampling artefacts. The model-based BLIP iterations improve this, however, due to lacking spatial regularisation, BLIP has limited accuracy and cannot fully remove aliasing artefacts (e.g. see T2 maps in Figure~\ref{fig:spiral_recon}) despite 20 iterations and very long runtime. In contrast, all deep learning methods outperform BLIP not only in accuracy but also in having 2 to 3 orders of magnitude faster reconstruction times\textemdash an important advantage of the learning-based methods. The proposed \proxnet\ consistently outperforms all baselines, including DM and learning-based methods, over all defined accuracy metrics. This is achieved due to learning an effective spatiotemporal model (only) for the proximal operator i.e. the $\G$ and \BLC\ networks, directly incorporating the physical acquisition model \eH\ into the recurrent iterations to avoid over-parameterisation of the overall inference model, as well as enforcing reconstructions to be consistent with the Bloch dynamics and the k-space data through the multi-term training loss~\eqref{eqs:loss}. The MRFCNN and SCQ over-parametrise the inference by 1 and 3 orders of magnitude larger model sizes (the SCQ requires larger memory than DM) and are unable to achieve \proxnet's accuracy e.g. see the corresponding over-smoothed T2 maps in Fig.~\ref{fig:spiral_recon}. Finally, we observe that despite having roughly the same model size (storage), the \textit{encoder alone} $\G$ predictions are not as accurate as the results of the \proxnet's recurrent iterations. By increasing the number of iterations $T$ we observe that the \proxnet's accuracy consistently improves despite having an acceptable longer inference time.

\section{Conclusions}
In this work we showed that the consistency of the computed quantitative maps with respect to the physical forward acquisition model and the Bloch dynamics is important for reliably solving the MRF inverse problem using compact deep neural networks.
For this, we proposed \proxnet, a learned model-based iterative reconstruction framework that directly incorporates the forward acquisition and Bloch dynamic models within a recurrent learning mechanism with a multi-term training loss. The \proxnet\ adopts a data-driven neural proximal model for spatiotemporal processing of the MRF data, TSMI de-aliasing and quantitative inference. A chief advantage of this model is its compactness (a small number of weights/biases to tune), which might makes it particularly suitable for supervised training using scarce quantitative MRI datasets. Through our numerical validations we showed that the proposed \proxnet\ achieves a superior quantitative inference accuracy, much smaller storage requirement, and a comparable runtime to the recent deep learning MRF baselines, while being much faster than the MRF fast dictionary matching schemes. In future work, we plan to evaluate the non-simulated scanner datasets with higher diversities and possible pathologies to further validate the method’s potential for clinical usage.

\section*{Acknowledgements}
%\noindent\textbf{Acknowledgements.}
We thank Pedro G\'{o}mez, Carolin Prikl and Marion Menzel from GE Healthcare for the quantitative anatomical maps dataset. DC and MD are supported by the ERC C-SENSE project (ERCADG-2015-694888).

%\vspace{-.4cm}

\bibliographystyle{splncs04}
\bibliography{reference}

\end{document}